\documentclass[10pt,twocolumn,letterpaper,]{article}
\usepackage[T1]{fontenc}
\usepackage[utf8]{inputenc}
\usepackage{authblk}
\usepackage{cvpr}              
\usepackage{pifont} 
\usepackage[dvipsnames]{xcolor}

\definecolor{cvprblue}{rgb}{0.21,0.49,0.74}
\usepackage[pagebackref,breaklinks,colorlinks,citecolor=cvprblue]{hyperref}
\usepackage{multirow}
\usepackage{graphicx}
\usepackage{flushend}
\usepackage{lipsum}
\usepackage{caption}
\usepackage{float}
\usepackage{balance}
\def\paperID{*****} 
\def\confName{CVPR}
\def\confYear{2024}

\author{
    {Yusheng Dai$^\dag$, Hang Chen$^\dag$, Jun Du$^\dag$\thanks{Corresponding author} , Ruoyu Wang$^\dag$, Shihao Chen$^\dag$, Jiefeng Ma$^\dag$, Haotian Wang$^\dag$,  Chin-Hui Lee$^\ddag$ \vspace{-0.6em}}\\
    {$^\dag$ University of Science and Technology of China, Hefei, China}\\
    { $^\ddag$Georgia Institute of Technology, Atlanta, America} \\
    {\tt\small jundu@ustc.edu.cn}
    \vspace{-1em}
}

\title{A Study of Dropout-Induced Modality Bias on Robustness to Missing Video Frames for Audio-Visual Speech Recognition}

\begin{document}

\setlength{\abovecaptionskip}{5pt}
\setlength{\textfloatsep}{5pt}
\setlength{\intextsep}{-5pt}
\vspace{-2em}
\maketitle

\begin{abstract}
Advanced Audio-Visual Speech Recognition (AVSR) systems have been observed to be sensitive to missing video frames, performing even worse than single-modality models. While applying the dropout technique to the video modality enhances robustness to missing frames, it simultaneously results in a performance loss when dealing with complete data input. In this paper, we investigate this contrasting phenomenon from the perspective of modality bias and reveal that an excessive modality bias on the audio caused by dropout is the underlying reason. Moreover, we present the Modality Bias Hypothesis (MBH) to systematically describe the relationship between modality bias and robustness against missing modality in multimodal systems. Building on these findings, we propose a novel Multimodal Distribution Approximation with Knowledge Distillation (MDA-KD) framework to reduce over-reliance on the audio modality and to maintain performance and robustness simultaneously. Finally, to address an entirely missing modality, we adopt adapters to dynamically switch decision strategies. The effectiveness of our proposed approach is evaluated and validated through a series of comprehensive experiments using the MISP2021 and MISP2022 datasets. Our code is available at \url{https://github.com/dalision/ModalBiasAVSR}.

\end{abstract}
\vspace{-10pt}
\section{Introduction}
\label{sec:introduction}

Audio-Visual Speech Recognition (AVSR) is a multimodal application inspired by human speech perception. It outperforms single-modality models, particularly in noisy environments, by introducing noise-invariant complementary information from visual cues. Pioneered by researchers, AVSR has achieved significant advancements across various benchmarks with a simple end-to-end design \cite{23:ma2021end, 4:pan2022leveraging}.


Recent research on AVSR focuses on more challenging real-life scenarios. Techniques such as reinforcement learning \cite{chen2023leveraging} and carefully designed fusion architecture \cite{xu2020discriminative,yu2020audio,hong2022visual} are used to accommodate varying noise levels and overlapping speech. Self-supervised learning \cite{haliassos2022jointly} and automatic labeling techniques \cite{ma2023auto} are applied facing insufficient audio-visual pairs and the absence of labels. To address the issue of aligning audio and visual modalities, various alignment modules have been developed \cite{sterpu2018attention,sterpu2020teach,hu2023cross}. However, restricted to the open-source datasets \cite{son2017lip,24:afouras2018deep,afouras2018lrs3}, most studies often assume that each video frame is recorded in relatively high quality, without blurring, corruption, or loss. There is growing evidence to suggest that current advanced AVSR systems are highly susceptible to perturbations in video modality \cite{hong2023watch, Dai2023icme}, resulting in significant performance degradation. In some cases, these systems even perform worse than single-modality models \cite{13:chang2022robustness,12:makino2019recurrent}.

\begin{figure}[!t]
    \centering
    \includegraphics[width=0.95\columnwidth]{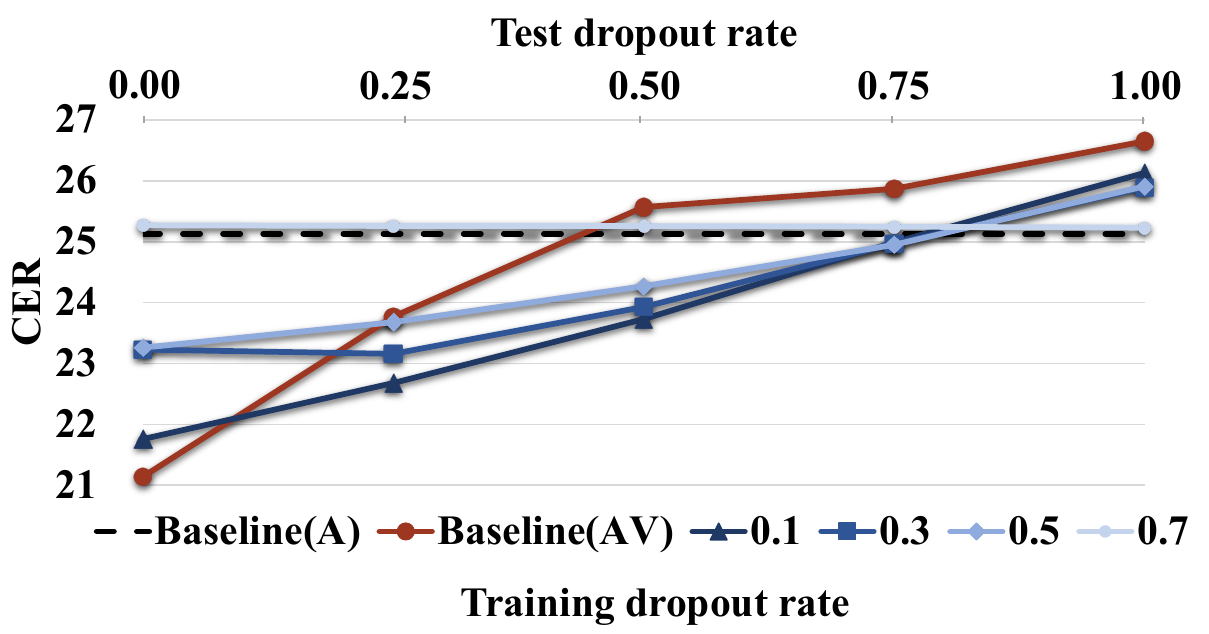}
    \caption{CER (in \%) degradation curves of AVSR trained with different dropout rates on video frames. Compared with the baseline AVSR without dropout (in red), other AVSR systems perform better with missing input but worse with complete data input. As the training dropout rate increases, the CER curve of AVSR gradually converges to that of ASR (dotted line).}
    \label{fig1:CER_Droprate}
\end{figure}

Missing video modality is a crucial and common problem for AVSR applied in real-life scenarios \cite{13:chang2022robustness,12:makino2019recurrent,son2017lip,hong2023watch}. It arises from various causes, including losses induced by network latency or hardware limitations, as well as errors in lip movement tracking due to occlusion of the lips and side-face poses. Most researchers use dropout techniques \footnote{Distinguished from the classic dropout in neural network training, dropout in this paper refers to a data augmentation technique that involves replacing original video frames with padding frames.} on training video data to improve the robustness against missing video frames \cite{13:chang2022robustness,12:makino2019recurrent,9:shi2022learning,8:hazarika2022analyzing,11:zhang2019robust}. And it has been proven to counter the out-of-distribution (OOD) issue resulting from missing modalities and mitigate performance degradation without additional inference consumption or complex modules. As illustrated in Figure~\ref{fig1:CER_Droprate}, a contradictory phenomenon could be observed:\label{contradictory phenomenon} \textit{while applying the dropout strategy to video training data can enhance the robustness against missing video modality, it also leads to performance degradation when dealing with complete data input. On the other hand, AVSR continues to lag behind ASR when facing completely missing video.}


 We attempt to analyze the reasons behind the above-mentioned phenomenon from the perspective of modality decision bias. Existing multimodal applications can be categorized into two types: (1) modality-balanced systems, in which each modality contributes relatively equally to the model decision, such as Multimodal Emotion Recognition (MER) \cite{2:zhao2021missing} and Hate Speech Detection (HSD) \cite{6:ma2022multimodal}; (2) modality-biased systems that over-relies on certain modality that contains more task-related information. AVSR is a typical modality-biased system dominated by audio. Therefore, an intuitive insight suggests that although dropout on the video modality could address the domain shift issue between the training and test set, it may exacerbate the modality decision bias on audio, subsequently demonstrating robustness towards missing video input.



In this paper, we first verify this intuitive speculation in Section \ref{sec:Analysis of dropout on visual modality} by quantitatively analyzing the differences between Mandarin AVSR and automatic speech recognition (ASR). The results uncover that the modality bias essentially represents a shift from a multimodal to a unimodal distribution on audio modality in latent representation subspace. Next in Section \ref{sec:MBH}, we extend our findings to partially modality-biased systems and propose a Modality Bias Hypothesis (MBH) to systematically describe the relationship between modality bias and robustness to missing modality. In Sections \ref{sec:MDA-KD} and \ref{sec:Adapter}, our aim is to enhance the robustness of AVSR without causing performance degradation with complete input and consistently outperform ASR model when faced with severe or complete video missing. To this end, we present Multimodal Distribution Approximation with Knowledge Distillation (MDA-KD), which leverages hidden knowledge extracted from complete data pairs to prevent task-relevant feature shifts towards a unimodal distribution. This approach facilitates the inference of incomplete modality inputs through the context information or the complete modality to acquire a modality-invariant representation. For video severely or entirely missing situations, adapters are adopted to the modality-specific branch to dynamically switch decision patterns based on modality-specific representations. We believe our key contributions can be summarized as follows:

$\bullet$ We investigate dropout-induced modality bias and uncover that it fundamentally manifests as a shift from a multimodal to a unimodal distribution of audio modality in the hidden representation subspace as detailed in Section ~\ref{sec:Analysis of dropout on visual modality}.

$\bullet$ We propose using the Modality Bias Hypothesis (MBH) to systematically describe the decision-making process influenced by modal bias in a multimodal system, as well as the relationship between modal bias and modality missing robustness as detailed in Section ~\ref{sec:MBH}.

$\bullet$ We propose Multimodal Distribution Approximation with Knowledge Distillation (MDA-KD) to enhance robustness against missing video and avoid performance degradation with complete input. For entirely missing modalities, adapters are adopted to dynamically switch decision bias to the specific modality as detailed in Section ~\ref{sec:Adapter}.


$\bullet$ We achieve top AVSR performances on MISP2021 and MISP2022 datasets while maintaining robustness against missing video frames as detailed in Section ~\ref{sec:experiment}.

\begin{figure}[!t]
    \centering
    \includegraphics[width=1.0\columnwidth]{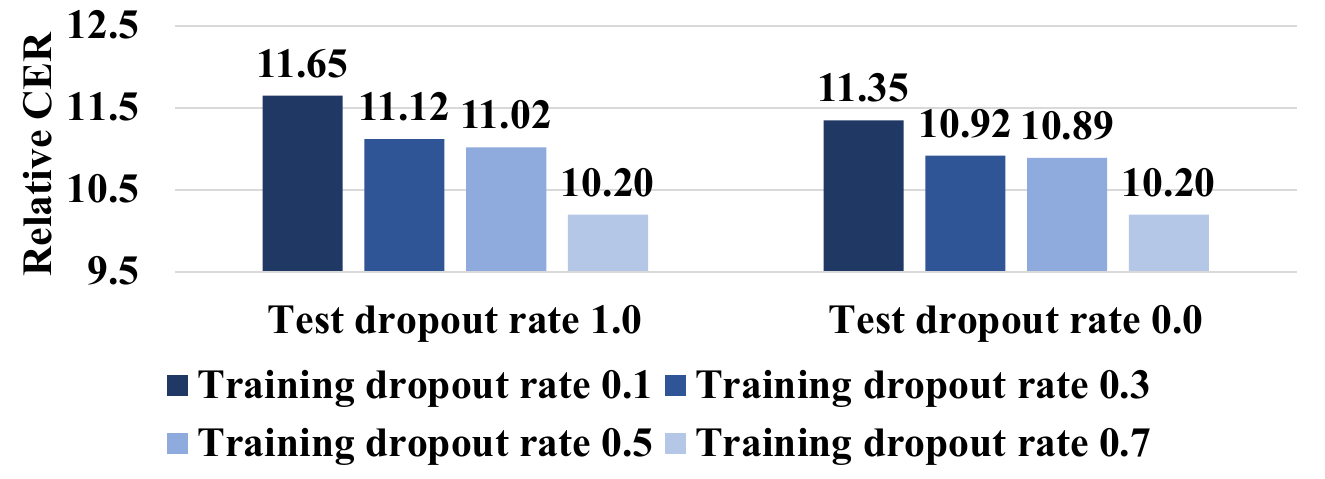}
    \caption{Two groups of similarity analysis between ASR and AVSR transcriptions. In both groups, an increase in the similarity of recognition transcriptions is observed as the training dropout rate increases. The similarity is measured by relative CER  (in \%), where the ASR transcription replaces the ground truth.}
    \label{fig2:Relative-CER_Droprate}
\end{figure}

\begin{figure}[!t]
    \centering
    \includegraphics[width=1.0\columnwidth]{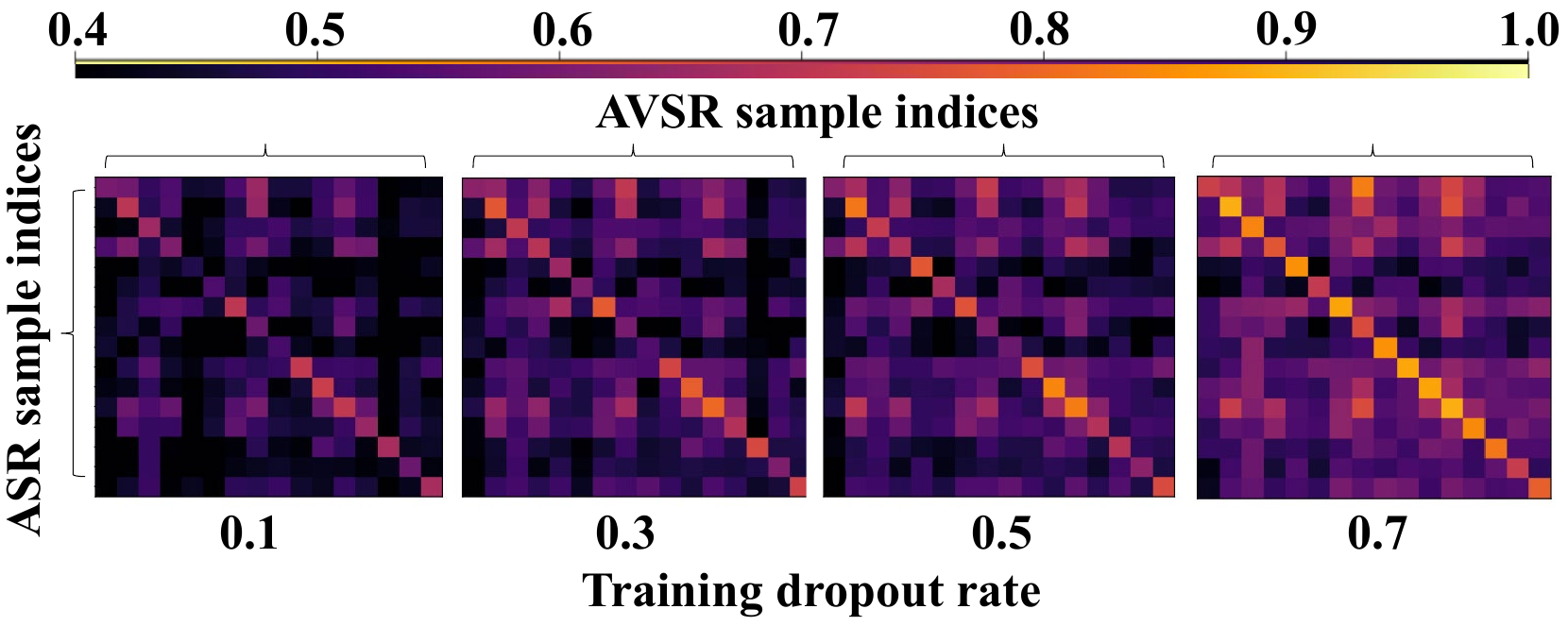}
    \caption{Similarity matrices of intermediate representations between ASR and different AVSR settings. As training dropout rates increase, the diagonal lines become brighter, indicating closer proximity between the multimodal and the unimodal distributions of the latent decisive subspace in AVSR.}
    \label{fig3:Training_Droprate}
\end{figure}

\begin{figure*}[!t]
\setlength{\belowcaptionskip}{-15pt} 
    \centering
    \includegraphics[width=1.9\columnwidth]{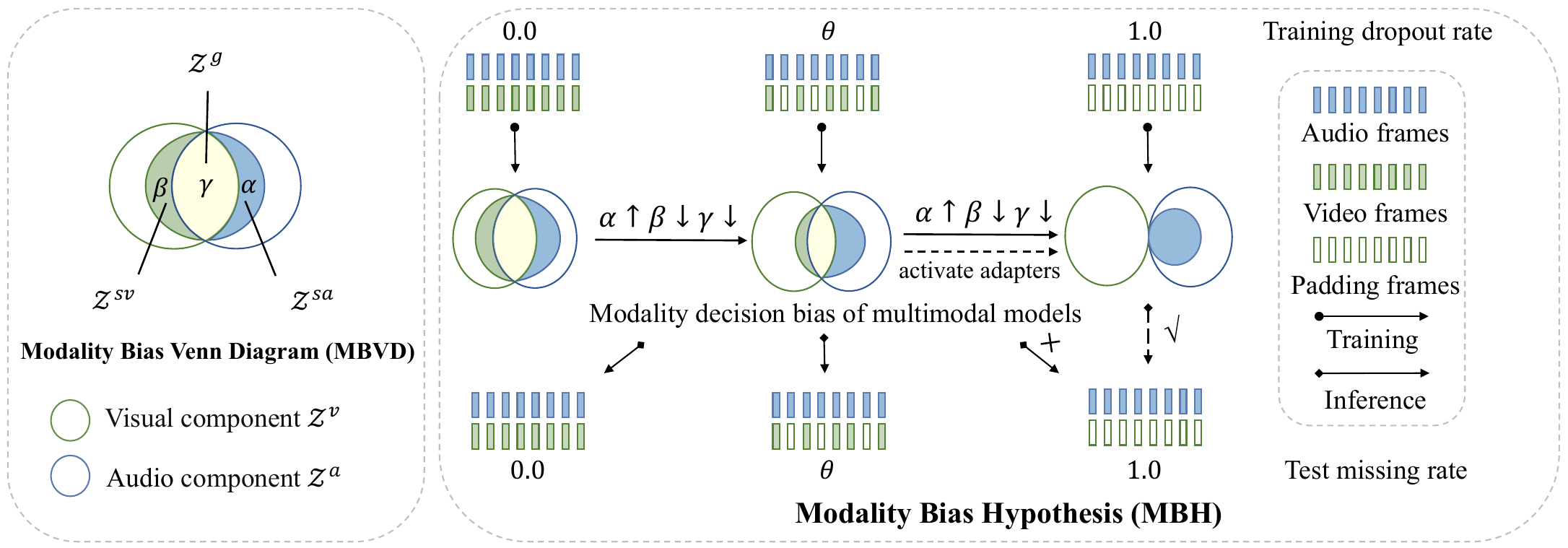}
    \caption{An illustration of the Modality Bias Hypothesis (MBH). In the left subplot, the task-relevant component (shaded part) of the latent representations consists of $Z^{sa}$, $Z^{sv}$ and $Z^g$, representing audio-specific, visual-specific decision features and modality-general decisive features respectively. The corresponding proportions are denoted by $\alpha$, $\beta$, and $\gamma$. The right subplot shows a dynamic process of decisive bias with an increasing training dropout rate. Dropout leads to a consistent modality bias on audio, regardless of the extent of the missing.}
    \label{fig4:MBH}
\end{figure*}

\section{Dropout-Induced Modality Bias}
\label{sec:Analysis of dropout on visual modality}

We investigate the contradictory phenomenon \ref{contradictory phenomenon} by examining the character error rate (CER) across five Mandarin AVSR systems varying training dropout rates (from 0.0 to 0.7) and testing video missing rates (from 0.0 to 1.0). As shown in Figure \ref{fig1:CER_Droprate}, two trends are observed: (1) in terms of absolute CER, the model trained with a higher dropout rate deteriorate more on no-missing complete multimodal data and slightly missing video frames, but it performs better on severely and entirely missing video frames; and (2) in term relative performance, the CER degradation curve of the AVSR model trained with a higher dropout rate tends to converge to the unimodal ASR recognition curve. We further ensure whether the similarity of performance degradation curves directly corresponds to the recognition transcription similarity of ASR and AVSR in Figure~\ref{fig2:Relative-CER_Droprate}. As we expected, an increase in training dropout rate leads to higher transcription similarity between AVSR and ASR across different test settings.

To understand this, we investigate the discrepancy in decisive patterns of ASR and each AVSR. We aim to quantify the divergence between latent decision distributions of these models by measuring the distance of intermediate representation samples. Through random sampling of complete audio-visual data batches, we generate intermediate layer representations using the encoder of ASR or AVSR trained at different dropout rates. Figure~\ref{fig3:Training_Droprate} illustrates cosine distance-based similarity matrices for the intermediate representations between ASR and different AVSR configurations. The diagonal elements in each subplot represent the similarity between intermediate representations from the same inputs. Notably, with an increase training dropout rate, these diagonal lines brighten, signifying a rise in intermediate representation similarity. This suggests closer proximity of the AVSR multimodal distribution in the latent decisive subspace to the unimodal distribution of ASR.

Through the aforementioned three experiments, we have discovered that increasing the training dropout rate on video data leads to increased similarity between AVSR and ASR in the performance degradation curves, recognition results, and intermediate representation subspace distribution. The findings reveal the significant impact of dropout in introducing effectively perturbs the distribution of multimodal training data. It leads to a shift from multimodal joint distribution to unimodal distribution, resulting in a decision bias towards audio during the decision-making process, as reflected in the output similarity of ASR. We refer to this phenomenon induced by dropout as dropout-induced modality bias. Although dropout-induced bias enhances the robustness of missing video data to some extent, we emphasize that it contradicts the primary design of AVSR as a robust application in noisy environments with supplementary visual cues. The introduction of artificial noise (padding frames) in video data induces the model to converge toward trivial solutions, leading to an excessive dependence on the audio modality. This over-reliance, in turn, leads to a degradation in performance when presented with complete multimodal input in a noisy environment.

\section{Modality Bias Hypothesis (MBH)}
\label{sec:MBH}
In this section, we propose the Modality Bias Hypothesis (MBH)  based on the Modality Bias Venn diagram (MBVD) to systematically describe the relationship between modality bias and robustness to missing modality.
\paragraph{Modality Bias Venn Diagram}
\vskip -10pt
As shown in Figure~\ref{fig4:MBH} on the left, the MBVD depicts the components of the latent decisive feature of multimodal systems in the form of a Venn Diagram. It is a variant of the Modality Venn Diagram (MVD) employed in multimodal knowledge distillation \cite{18:xue2022modality}. Without loss of generality, we take AVSR as an example and define $\mathcal{X}^{a}$, $\mathcal{X}^{v}$, and $\mathcal{Y}$ as the original feature space of audio, video and label space, respectively. The decisive feature $z$, commonly a form of intermediate layer representation, consists of two modality components $z^a$ (blue circles) and $z^v$ (green circle). We denote $I(\cdot)$ as mutual information and $I(\cdot|\cdot)$ as conditional mutual information. The task-relevant decisive feature $z^u$ ($I(z,y)$) is depicted by the shaded region and can be further divided into three components. $z^g$ ($I(z^a,z^v,y)$) represents modality-general decisive features, while $z^{sa}$ ($I(z^u,z^a|z^g)$) and $z^{sv}$ ($I(z^u,z^v|z^g)$) represent modality-specific decisive features. We denote their proportions in $z^u$ as $\alpha$, $\beta$, and $\gamma$, respectively. These features collectively contribute to determining the final task output $\hat{y}$. For AVSR, a higher $\alpha$ represents a greater decision bias of the model on the audio modality, focusing more on speech than lip movements. A larger $\gamma$ indicates a model's inclination towards modality synergy by maximizing the mutual information between modalities for decision-making, as in some modality-balanced models ~\cite{2:zhao2021missing,6:ma2022multimodal}. Furthermore, $z^u$ is generated by the original features $x^{a},x^{v}$ as $g\left(x^a, x^v ; \phi\right)$, where $g(\phi)$ can be seen as a neural network-based transfer such as an encoder with parameters $\phi$. Therefore, the decision process of the multimodal system can be decomposed into two steps, following the Bayesian process: the MBVD hidden decisive feature generation step and the decision step:
\begin{equation}
P\left(y \mid x^a, x^v\right)=P\left(y \mid z^\mu\right) P\left(z^\mu \mid x^a, x^v\right)
\end{equation}

\paragraph{Modality Bias Hypothesis} 
\label{MBH}
\vskip -5pt
Based on MBVD, we give a systematic description of the relationship between modality bias and robustness to missing modality in the view of MBH. As shown in Figure \ref{fig4:MBH} on the right, by applying dropout with different rates $k_i \in [0,1]$ on video training data, the original video feature space $\mathcal{X}^{v}$ can be split into a series of subsets $\{\mathcal{X}^{v}_{k_1},\mathcal{X}^{v}_{k_2},...,\mathcal{X}^{v}_{k_n}\}$. The samples from space $\mathcal{X}^{a} \times \mathcal{X}^{v}_{k_i}$ are denoted as dyads $\left(x^{a},x^{v}_{k_i}\right)$. Compared to the model trained on complete multimodal datas $\left(x^{a},x^{v}_{0.0}\right)$, the model trained on data pairs $\left(x^{a},x^{v}_{\theta}\right)$ with a video dropout rate $\theta_{train} \in (0.0,1.0)$ exhibits a greater decision bias on audio modality with larger $\alpha$, smaller $\beta$, and $\gamma$. As $\theta$ approaches 1.0, the task-relevant decisive feature $z_u$ becomes steadily dominated by the audio-specific decisive feature $z_a$, resulting in a transformation from a bimodal distribution in the latent representation subspace to a unimodal one. The decision pattern of the multimodal model shifts from $p(y|z_u)$ to $p(y|z_a)$. 

During the inference stage, these multimodal models display different modality decision biases. For the model trained on complete multimodal data or dropout on audio with a larger $\gamma$, they tend to search general information shared among modalities. This hypothesis effectively explains the observed experimental phenomena in previous studies. For modality-biased models, such as Multimodal Sentiment Analysis (MSA) \cite{8:hazarika2022analyzing} dominated by text, Multimodal Speech Enhancement (MSE) \cite{chen2021correlating} dominated by audio, as well as AVSR dominated by audio \cite{9:shi2022learning,10:zhou2019modality,11:zhang2019robust}, it has been observed that applying dropout on the primary modality helps alleviate modality bias and brings about slight improvements when dealing with complete input. On the other hand, the AVSR model with larger $\alpha$ and smaller $\gamma$ values tends to focus more on speech and neglect complementary information from lip movements. When dealing with partially or completely missing video data, the model with larger $\alpha$ shows its robustness, which aligns well with the aforementioned experimental observations.

\section{Multimodal Distribution Approximation with Knowledge Distillation (MDA-KD)}
\label{sec:MDA-KD}

\begin{figure*}[!t]
    \centering
    \setlength{\belowcaptionskip}{-15pt} 
    \hspace{-2mm}
    \includegraphics[width=1.9\columnwidth]{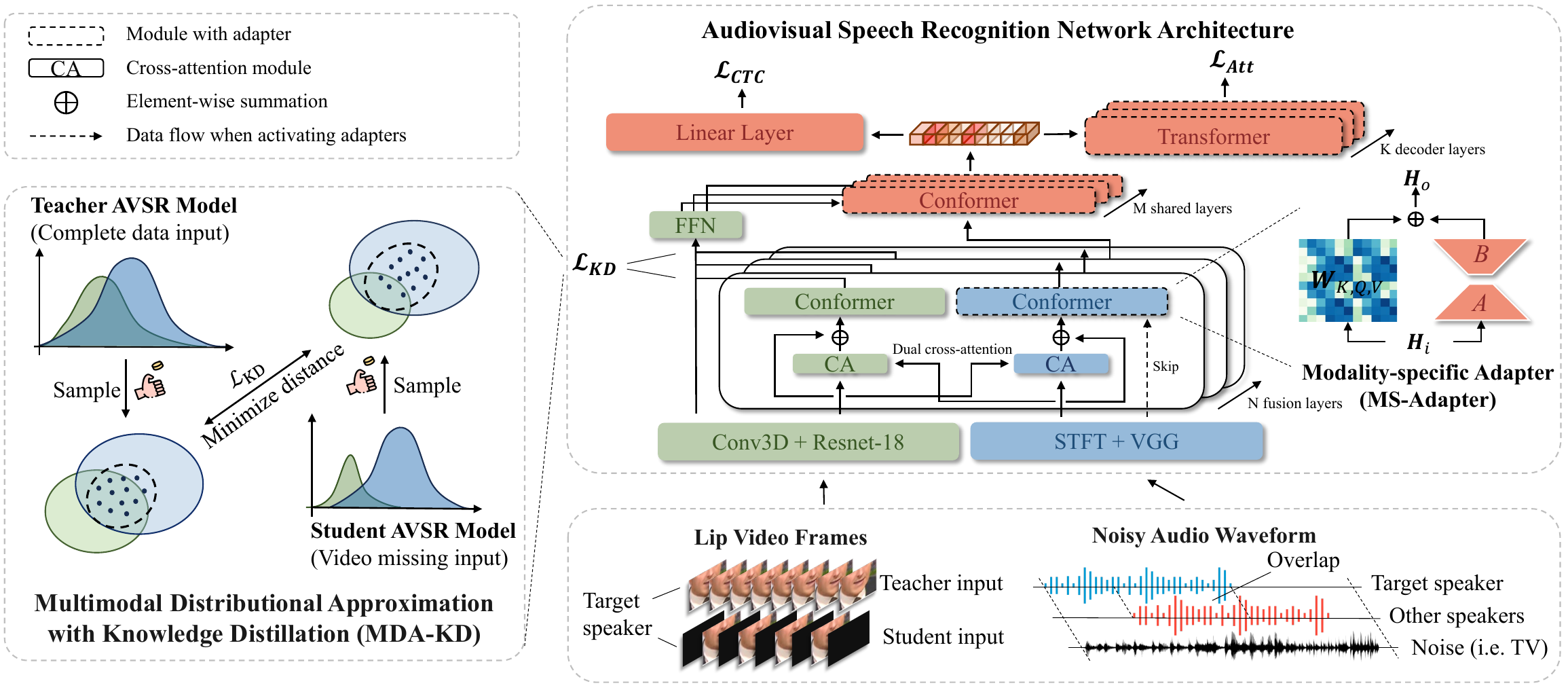}
    \caption{Overall framework of the proposed AVSR system. We address challenging real-world scenarios involving missing video frames and noisy speech with an overlap rate exceeding 40\% during both the training and testing stages. In MDA-KD, latent knowledge is sampled from the latent distribution of the teacher model with complete data input. This latent konwledege serves as an anchor point to prevent dropout-induced modality bias during the robustness training of the student network. For entirely missing video input, the MS-Adapter is activated to enable a dynamic decision switch.}
    \label{fig5:MDA-KD.}
\end{figure*}

For the robustness training of modality-bias systems, it is crucial to avoid dropout-induced modality bias on the primary modality. Dropout indeed alleviates the OOD problem to some extent but encourages multimodal models to pursue trivial solutions at the same time. Ideal robust multimodal models are expected to achieve two goals: (1) learn to extract mutual information across modalities rather than relying on a certain modality when facing complete paired input, and (2) learn to complement information from the other modality and utilize context information from adjacent frames. To prevent excessive modality bias caused by dropouts, we propose a novel Multimodal Distribution Approximation with Knowledge Distillation (MDA-KD) framework to constrain the distribution of the multimodal feature space during the robustness training phase. 

Unlike traditional knowledge distillation methods, firstly, the teacher model is trained on the complete multimodal data pairs, while the student model is trained on missing video data. The teacher model is relatively unbiased with a higher proportion of modality-general decisive features $z^g$ in the MBVD space. During the training process of the student model, the teacher model serves as an anchor point, preventing the student model from shifting towards a unimodal distribution in the audio modality. Note that the difference between teacher and student models in our method is modality bias varies, rather than size, architecture as in common KD methods \cite{chen2022dearkd,valverde2021there,xue2021multimodal,peng2019correlation}. Additionally, distillation occurs at the hidden layer rather than the logistic outputs, aiming to minimize the distances between decision distribution samples of the teacher and student models and constrain the intermediate representation subspace distribution of the student model. In practice, we take the knowledge from the intermediate representation of the cross-modal encoder layers.

Here, we adopt the symbol definitions from Section \ref{MBH} and provide a formal description of MDA-KD. For a naturally modal-biased multimodal system, the data samples from original feature space $\mathcal{X}^{a} \times \mathcal{X}^{v}_{k_i} \times \mathcal{Y}$ can be denoted as triples $\left(x^{a},x^{v}_{k_i}, y\right)$. For simplicity, we denote $x^{v}_{0.0}$ as $x^{v}$. The teacher model $Te(\phi)$ is first trained on complete multimodal data $\left(x^{a},x^{v}, y\right)$ model with parameters $\phi$, and the model's decision process can be formulated as $P_{te}\left(y \mid x^a, x^v\right)$ in a Bayesian decision problem. We assume that the teacher model is a neural network $g\left(\phi\right)$ and it is trained by minimizing the following loss function, a form of multitask learning.

\begin{equation}
Te(\phi)=\min _\phi \mathcal{L}_{\text{MLT}}\left(g\left(x^a, x^v ; \phi\right), y\right),
\end{equation}
\vspace{-5pt}
\begin{equation}
\begin{aligned}
\label{loss}
\mathcal{L}_{\text{MLT}}(x^a, x^v; \phi) &= \lambda \log P_{\text{CTC}}(y \mid x^a, x^v) \\
&\quad + (1-\lambda) \log P_{\text{Att}}(y_i \mid x^a, x^v),
\end{aligned}
\end{equation}

\noindent where the tunable parameter $\lambda \in [0,1]$ is used to balance the sequence-level Connectionist Temporal Classification (CTC) loss and the frame-wise Cross Entropy (CE) loss, which serve as the standard end-to-end ASR training objectives.
During the training of the student model, the dropout strategy is applied to the secondary modality $v$, while the teacher model is frozen with complete multimodal data as input. It is important to note that the student and teacher models have the same network architecture. From the perspective of MBVD, the whole decision process of the multimodal model can be divided into hidden feature generation step and decision step.

\begin{equation}
P_{st}\left(y \mid x^a, x^v_{k_i}\right)=P_{st}\left(y \mid z^\mu\right) P_{st}\left(z^\mu \mid x^a, x^v_{k_i}\right),
\label{zu1}
\end{equation}
\begin{equation}
P_{te}\left(y \mid x^a, x^v\right)=P_{te}\left(y \mid z^\mu\right) P_{te}\left(z^\mu \mid x^a, x^v\right),
\label{zu2}
\end{equation}

where $z^\mu\in \mathbb{R}^{d \mu}$ represents the combined representation of modality-specific decisive features $z^{s a} \in \mathbb{R}^{d a}$, $z^{s v} \in \mathbb{R}^{d v}$, and modality-general decisive features $z^g \in \mathbb{R}^{d g}$. The tuple $\left(z^{s a}, z^{s v}, z^g \right)$ represents a sample drawn from the MBVD hidden features space, denoted as $\mathcal{Z}^{s a} \times \mathcal{Z}^{s v} \times \mathcal{Z}^g$.

Initialized on the parameter of the teacher model, we introduce an additional loss term to constrain the dynamic process of the student model's MBVD feature distribution in robust training. The distance between batch samples from the student and the teacher model is used to approximate the difference of distribution, which serves as a form of frame-level knowledge distillation.
\begin{equation}
\begin{aligned}
\mathcal{L}_{\text{KD}}&(x^a, x^v, x^v_{k}; \phi_{te}, \phi_{st}) = \text{KL}\left(
    S_{te}, S_{st}
    \right), \\
S_{te} &= \sigma_T\bigl(\operatorname{Sample}(P_{te}(z^\mu \mid x^a, x^v))\bigr), \\
S_{st} &= \sigma_T\bigl(\operatorname{Sample}(P_{st}(z^\mu \mid x^a, x^{v }_{k_i}))\bigr),
\end{aligned}
\end{equation}
\label{teacher}
\noindent where $\sigma_T(x)$ denotes the SoftMax function with temperature $T$ and $Sample$ represents the sample function. This distribution approximation serves two main purposes. Firstly, during training, when the student network encounters a missing modality feature $x^v_{k_i}$, the convergence of the student's decisive feature $z^u=g(x^a,x^v_{k_i};\phi_{st})$ towards the teacher's decisive feature $z^u=g(x^a,x^v;\phi_{te})$ encourages the utilization of contextual information from $x^v_{k_i}$. Additionally, with the dual cross-attention design, the process complements the information extracted from $x^a$, effectively addressing the condition of missing frames and promoting out-of-distribution generality. On the other hand, the KD loss is used to minimize the distance between the distributions of the teacher and student models, preventing the student model from converging to trivial solutions. Subsequently, we train the student model jointly with a weighted sum of the standard training loss and distillation loss:

\begin{multline}
\mathcal{L}_{\text{MLT}}(x^a, x^v, x^v_{k}; \phi_{te}, \phi_{st}) = \beta \mathcal{L}_{\text{KD}}(x^a, x^v, x^v_{k}; \phi_{te}, \phi_{st}) \\
+ (1-\beta) \mathcal{L}_{\text{MLT}}(x^a, x^v_k; \phi_{st}).
\end{multline}

\section{Modality-Specific Adapter (MS-Adapter)}
\label{sec:Adapter}
As illustrated in Figure \ref{fig4:MBH} on the right, when facing severely or entirely missing video data, we consider it unreliable to continue employing a synergistic decision-making strategy like MDA-KD with relatively high values of $\gamma$ and $\beta$. Padding frames lack sufficient contextual information and may introduce noise. Therefore, in such scenarios, a dynamic switch in decision strategy from $P(y|z^u)$ to $P(y|z^a)$ is necessary as a complement to MDA-KD. In view of the success of adapters applied in foundation model fine-tuning \cite{zhang2021tip,houlsby2019parameter,rebuffi2017learning,gao2023clip}, we attempt to extend it to address the modality missing issue in multimodal models. For clarity, we refer to this extension as Modality-Specific Adapter (MS-Adapter). Specifically, LORA \cite{hu2021lora} is adopted to self-attention layers in the audio branch, marked with a dashed box in Figure \ref{fig5:MDA-KD.}. These adapters perform residual-style feature blending with the original pre-trained features. The residual weight could be represented as low-rank matrices $\Delta W\in \mathbb{R}^{d \times d }$, and it could be decomposed into a pair of fan-in and fan-out linear layers with weights $A\in \mathbb{R}^{r \times d }$ and $B\in \mathbb{R}^{d \times r }$ ( $r\ll d$ ). The reparametrization operation can be formulated below. 
\begin{equation}
H_o=H_i(W_0+ \Delta W) = H_i(W_0+B A)
\end{equation}

By activating the MS-Adapter, we can dynamically switch the decision-making pattern by activating the adapters. We highlight two advantages of the MS-Adapter. First, a substantial amount of unpaired unimodal training data and data augmentation techniques could be used in the training process of the adapters. Second, the adapter training process provides an opportunity to modify the computation pathway. As illustrated in Figure~\ref{fig5:MDA-KD.} with dashed arrows, in both training and inference stage with audio-only input, the computation flow of the video branch will be directly cut off, and the modality fusion cross-attention module will be skipped to reduce computational costs.


\begin{table*}[]
\centering
\setlength{\belowcaptionskip}{-15pt} 
\begin{tabular}{l|ccccc|ccccc}
\toprule[1pt]
\multirow{2}{*}{\textbf{Model}} & \multicolumn{5}{c|}{\textbf{Training setttings}}                             & \multicolumn{5}{c}{\textbf{Test dropout rate}}                                   \\
                                   & \textbf{Dropout} & \textbf{\boldmath{D$_{prob}$}} & \textbf{Init.} & \textbf{MDA-KD} & \textbf{MS-Adapter} & \textbf{0.00}  & \textbf{0.25}  & \textbf{0.50}  & \textbf{0.75}  & \textbf{1.00} \\ \hline
A0                                 & \text{\ding{55}}              & 0.0   & Random   & \text{\ding{55}} & \text{\ding{55}}     & 25.13          & 25.13          & 25.13          & 25.13          & 25.13                                     \\
AV0                              & \ding{55}               & 0.0 & A0  & \text{\ding{55}}              & \text{\ding{55}}               & 21.14          & 23.77          & 25.57          & 25.87          & 26.65                                      \\ \hline
AV1                                & \ding{52}             & 1.0   & A0              & \text{\ding{55}}              & \text{\ding{55}}                & 23.26          & 23.68          & 24.27          & 24.95          & 25.91                                       \\
AV2                                & \ding{52}             & 0.5 & A0                & \text{\ding{55}}              & \text{\ding{55}}               & 21.72          & 22.56          & 23.37          & 24.46          & 25.64                                      \\
AV3                               & \ding{52}            & 0.5  & AV0                & \text{\ding{55}}             & \text{\ding{55}}               & 21.53          & 22.47          & 23.65          & 24.55          & 25.90                                      \\
AV4                                 & \ding{52}           & 0.5     & AV0            & \ding{52}\rotatebox[origin=c]{-9.2}{\kern-0.7em\ding{55}}          & \text{\ding{55}}    & 21.38 & 22.18 & 23.20 & 24.40 &25.70                                          \\
AV5                                 & \ding{52}           & 0.5     & AV0            & \ding{52}               & \text{\ding{55}}               & 21.11          & 21.77          & 22.78          & 24.02          & 25.45                                         \\

AV6                                & \ding{52}            & 0.5    & AV0             & \ding{52}                & \ding{52}                 & \textbf{21.11} & \textbf{21.77} & \textbf{22.78} & \textbf{24.02} & \textbf{24.94}   \\ \bottomrule[1pt]
\end{tabular}
\caption{An overall comparison in CER (\%) of different system configurations. Different from the dropout rate, $D_{prob}$ represents the proportion of data with missing frames in the training set. Init. refers to the network initialization method.}
 \label{table:ablation study}
\end{table*}

\section{Experiment Settings}
\label{sec:Experiment setup}
\paragraph{Dataset}
\label{subsec:Dataset}

We conduct experiments using two open-source datasets: MISP2021 \cite{9746683} and MISP2022 \cite{wang2023multimodal}. MISP2021 presents a large-scale audio-visual corpus recorded in real-life home TV scenarios with multiple groups of speakers chatting simultaneously. Multiple microphone arrays and cameras are used to collect far/middle/near-field audio and far/middle-field videos. MISP2022 further elevates the difficulty by introducing session-level AVSR tasks without clear speech segmentation. Moreover, a test set with more missing video frames is also included. Compared to datasets like LRS3 \cite{afouras2018lrs3}, MISP provides real-world audio-visual data with some blurriness and high missing rates in video frames, alongside various background noises and speech overlaps (about 42\% in the training set and 49\% in the test set). All experiments were exclusively evaluated using far-field data to ensure a technique's generalizability, avoiding the need for simulating specific noise types.

\paragraph{Implementation Detail} \vskip -10pt
We strictly adhere to the approaches outlined in \cite{Dai2023icme} for model training and network architectures. We initialize the AVSR model with two pre-trained unimodal models and fine-tune it in an end-to-end manner. As shown in Figure~\ref{fig5:MDA-KD.}, the AVSR model follows a dual-branch network where $N=3$, $M = 9$ and $K = 6$. For the loss function, we set $\lambda$ to 0.7 in Equation \ref{loss} and CTC loss consists of the same weighted intermediate CTC \cite{lee2021intermediate} losses in layers [3, 6, 9, 12]. In Equation \ref{teacher}, we use 0.1 for $\beta$.~\ref{shared}

\paragraph{Dropout Settings} \vskip -10pt
\label{Dropout out settings}

Similar to \cite{13:chang2022robustness}, we evaluate the robustness to missing video modality using various dropout methods and rates: Segment Dropout, Utterance Dropout, and Interval Dropout. Testing involves dropout rates from 0.0 to 1.0 in 0.25 intervals, and results from three dropout methods are averaged at each rate to obtain overall dropout results. When conducting ablation studies, segments with naturally missing video frames (17\%) are systematically excluded from the test set, ensuring a consistent and controlled video missing rate. In our method, during training, each sample is assigned a random dropout method from the three methods and an extra one from~\cite{9:shi2022learning} with an optimized dropout rate. Training and testing stages involve filling missing video frame pixels with zeros instead of using interpolation or repetition. We conducted a hyper-parameter search over the training dropout rate and found that 0.5 is optimal for our method. This rate implies that half of the video frames are padded with zero in a selected segment.~\ref{shared}
\footnotetext[3]{\label{shared}More details can be found in Appendix.}


\section{Experiments and Result Analysis}
\label{sec:experiment}

\begin{table}[!t]
\centering
\begin{tabular}{lclcl}
\toprule[1pt]
\textbf{Insert part} & \textbf{Rank} & \textbf{DA} & \textbf{Params(MB)} & \textbf{CER(\%)} \\  \hline
Encoder              & 32         & \ding{55}   & 4.50      & 25.35            \\
Encoder              & 32         & \ding{52}   & 4.50       & 25.08            \\
En\&Decoder          & 32         & \ding{52}   & 9.00       & 25.20            \\
Encoder              & 64         & \ding{52}   & 9.00       & 25.08            \\
En\&Decoder          & 64         & \ding{52}   & 18.00      & 25.05            \\
Encoder              & 128        & \ding{52}   & 18.00      & 25.01            \\
En\&Decoder          & 128        & \ding{52}   & 36.00     & \textbf{24.94   }         \\ \bottomrule[1pt]
\end{tabular}
\caption{Performance analysis of MS-Adapter. DA means data augmentation, including speed perturbation and utterance concat.}
\label{table3:Model Rank Insert part CER}
\end{table}

\subsection{Overall Comparison of Experiment Settings}
In Table \ref{table:ablation study}, we conduct an ablation study to validate the impact of the proposed method on the MISP2022 dataset with oracle speech segmentation. In this investigation, we establish two baselines following \cite{Dai2023icme} trained on complete modality data, referred to as A0 and AV0. AV0 is fine-tuned based on A0 and a pre-trained ResNet-18 visual extractor. We commence by examining the impact of dropout probability $D_{prob}$ in the video training data. In contrast to AV1, AV2 introduces half of the complete data pairs. Consequently, it mitigates dropout-induced modality bias and prevents performance degradation on complete data input to some extent. A higher proportion of complete data encourages the model to utilize general information across modalities. This finding aligns with previous research ~\cite{13:chang2022robustness}, highlighting the superiority of utterance dropout over random frame dropout (the former means a larger $D_{prob}$). Compared to AV2, AV3 is fine-tuned from AV0, starting within a relatively stable convergence range with complete modality input. However, during fine-tuning, the distribution of new incomplete data pairs tends to disrupt the balanced state, whereas the distribution of complete data pairs exhibits the opposite effect. Thus, AV3 outperforms AV2 with lower test missing rates but lags when facing severe video absence, depicting a tug-of-war dynamic without clear guidance.

Next, we validate the effectiveness of MDA-KD. When comparing AV3 to AV5, the latter demonstrates superior performance for both complete and missing video modality inputs. AV4 successfully achieves our goal of enhancing robustness without any performance degradation on complete input (21.11\% vs. 21.14\%). This implies that the teacher model AV0 provides an explicitly optimized target in the robustness training. It effectively constrains the distribution shift to the audio modality, preventing excessive modality bias caused by dropout. Furthermore, in AV4, we restrict the flow of audio data into the video branch in the dual cross-attention module during both the training and testing stages. Consequently, a performance drop is observed in all test suites compared with AV5, indicating that our method leverages modality-general information from audio modality to complement the missing information. Subsequently, we integrate MS-Adapters into the audio branch of the final proposed model AV6 building upon AV5. Consequently, the performance with audio-only input improves to a 24.94\% CER, surpassing A0 for the first time (24.94\% vs. 25.13\%). These results show the effectiveness of MS-Adapters by dynamically switching to the decision patterns on audio modality with audio-only input.

\begin{table}[!t]
\centering
\resizebox{1.0\columnwidth}{!}{
\begin{tabular}{lccccc}
\toprule[1.2pt]
\multirow{2}{*}{\textbf{Method}} & \multicolumn{5}{c}{\textbf{Test dropout rate}} \\
                                 & \textbf{0.00}  & \textbf{0.25}  & \textbf{0.50}  & \textbf{0.75}  & \textbf{1.00}  \\ \hline
Cascade Utt \cite{13:chang2022robustness}                     & 22.54 & 23.89 & 25.23 & 26.05 & 28.15 \\
AV Dropout Utt  \cite{9:shi2022learning}         &  22.00 & 23.37 & 25.35 & 26.21 & 26.78 \\
Dropout Utt \cite{12:makino2019recurrent}                     & 22.08 & 23.21 & 24.56 & 25.08 & 25.46 \\ 
\textbf{Ours}                    & \textbf{21.11} & \textbf{21.77} & \textbf{22.78} & \textbf{24.02} & \textbf{24.94} \\ \bottomrule[1.2pt]
\end{tabular}
}
\caption{A CER(\%) comparison with other dropout methods.}

\label{table2:drpout comparison}
\end{table}


\begin{table*}[]
\centering
\setlength{\belowcaptionskip}{-15pt} 
\begin{tabular}{lllllcc}
\toprule[1pt]
\multirow{2}{*}{\textbf{Benchmark}} & \multirow{2}{*}{\textbf{System}} & \multicolumn{2}{c}{\textbf{Training Data}}      & \multirow{2}{*}{\textbf{Backbone}} & \multirow{2}{*}{\textbf{Obj. Function}} & \multirow{2}{*}{\textbf{CER / cpCER(\%)}} \\
                                    &                                  & \multicolumn{1}{c}{A}          & \multicolumn{1}{c}{V}                                  &                                    &                                     &                                         \\ \hline
\multirow{4}{*}{MISP2021}           
                                    & SJTU \cite{wang2022sjtu}                             & 300 hours  & \multicolumn{1}{l}{LRW-1000}       & Conformer                          & ED + SE                               & 34.02                                   \\
                                    & NIO \cite{xu2022channel}                                 & 3300 hours & \multicolumn{1}{l}{LRW-1000 \cite{yang2019lrw}}       & Transformer                        & ED                                  & 25.07                                   \\
                                    & USTC \cite{Dai2023icme}                             & 500 hours  & \multicolumn{1}{l}{w/o extra data} & Conformer                          & ED                                  & 24.58                                   \\
                                    & \textbf{Ours}                    & 1000 hours & w/o extra data & Conformer                        & ED + InterCTC                         & \textbf{21.53}                          \\ \hline
\multirow{4}{*}{MISP2022}           
                                    & NIO \cite{nio2022}                              & 3300 hours & LRW-1000                           & Conformer                          & ED                                  & 29.58                                   \\
                                    & XMU \cite{li2023xmu}                             & 2100 hours & LRW-1000                           & Conformer                          & ED + InterCTC                         & 31.88                                   \\
                                    & NPU \cite{wang2024mlca}                                & 1300 hours & w/o extra data                           & E-Branchformer                          & ED + InterCTC                                  & 29.13                                 \\
                                    & \textbf{Ours}                    & 1000 hours & w/o extra data                     & Conformer                          & ED + InterCTC                         & \textbf{28.06}
\\ \bottomrule[1pt]
\end{tabular}

\caption{A Comparison of training data, architecture, Objective function, and overall performance with state-of-the-art systems in MISP2021 and MISP2022. InterCTC refers to Intermediate CTC loss \cite{lee2021intermediate}, the ED loss is formulated in Equation~(\ref{loss}) and SE represents the mean square error loss commonly used in speech enhancement \cite{wang2022sjtu}. For the session-level AVSR task on the MISP2022 dataset, we use evaluate the performance using the concatenated minimum-permutation character error rate (cpCER) \cite{watanabe2020chime} metric.}
\label{table4:SOTAComparison}
\end{table*}

\subsection{Validation of MS-Adapter}
We further explore three key factors in MS-Adapter adaptation: data augmentation, insert part and rank dimension. In Table \ref{table3:Model Rank Insert part CER}, we observe a decrease in CER from 25.45\% (AV4) to 25.35\%, and it further improves to 25.08\% with data augmentation doubling audio training data. These results suggest that the adapter adaptation effectively enhances the robustness of AVSR with completely missing video, requiring only an additional 4.50MB in parameters. It provides an opportunity to apply data augmentation that is effective for unimodal model training and to use extra unpaired data. Next, increasing the ranks and the quantity of adapters results in further performance gains at the expense of a larger parameter. The best performance, achieving 24.94\%, is shown in the bottom row and attained with the adapter inserted in both encoder and decoder blocks.

\subsection{Comparisons with Other Dropout Techniques}
As shown in Table ~\ref{table2:drpout comparison}, we compare our proposed framework with three widely used dropout techniques \cite{13:chang2022robustness,9:shi2022learning,12:makino2019recurrent}. Cascade Utt employs a separable cascade structure that an AV model is superimposed on an audio-only model. Inputs are then directed through either the audio-only or the AV path with a probability $p_1$. AV Dropout Utt randomly drops either the entire video or the entire audio segments, each with a probability $p_2$. Dropout Utt exclusively drops the video segments with a probability $p_3$. We adopt the optimal dropout settings from \cite{13:chang2022robustness}, where $p_1 = 0.25$, $p_2 = 0.25$, and $p_3 = 0.5$. For Cascade Utt, we follow \cite{13:chang2022robustness} to build the network and maintain comparable parameters with ours. As a result, Cascade Utt maintains relative robustness with consistent improvement trends but performs poorly in absolute performance. Clearly, our improved dropout strategy outperforms the other three techniques in all test suites and does not cause performance degradation.

\subsection{Comparisons with State-of-the-art Systems}
Finally, we compare our system with the state-of-the-art systems on the MISP2021 and MISP2022 challenges\cite{xu2022channel,wang2022sjtu,Dai2023icme,nio2022,guo2023npu,li2023xmu} as shown in Table~\ref{table4:SOTAComparison}. With Recognizer Output Voting Error Reduction (ROVER) \cite{fiscus1997post}, we rescore the output transcripts of A0, AV0, and A6 mentioned in Table~\ref{table:ablation study}. In the MISP2021 utterance-level AVSR challenge with oracle speech segmentation, our system outperforms the previous SOTA system by achieving an absolute CER reduction of 3.05\% from 24.58\% to 21.53\%. Our top-performing system, AV5, attains a CER of 22.13\%. Moving to the MISP2022 session-level AVSR challenge, we build our diarization system closely adhering to \cite{cheng2023whu}. We secure a ROVER cpCER score of 28.06\% and obtain the best system score with a cpCER of 28.55\%.  When oracle segmentations are utilized, our system achieves a ROVER CER score of 21.80\% and the best model score of 21.53\% in CER.

\vspace{-2pt}
\section{Related Works}
\label{sec:Related Work}

\paragraph{Modality Missing in Multimodal Learning} 
\label{subsec:Missing Modality}
The prevalent issue of missing modalities in multimodal applications has prompted research that specifically targets severe modality 
absences. Generative models \cite{19:suo2019metric,20:cai2018deep} and meta-learning predict missing modalities using available or few-shot paired samples. Balanced models utilize joint multimodal representations \cite{21:wang2020transmodality,22:pham2019found,li2023mseg3d}. Models addressing modality bias employ data augmentation methods like modality dropout \cite{8:hazarika2022analyzing,13:chang2022robustness} to tackle out-of-distribution challenges. For AVSR, we prioritize efficiency and opt for dropout due to its plug-and-play nature and lightweight implementation. More discussion could be found in Appendix.
\vspace{-2pt}
\paragraph{Video Modality Robustness in AVSR} \vskip -10pt
\label{subsec:Missing Modality in AVSR}
To enhance performance on low-resolution videos, visual extractors are commonly pre-trained on relatively high-quality videos with isolated words \cite{23:ma2021end} or acoustic pseudo-labeling classification tasks \cite{Dai2023icme}. Addressing situations involving corruption, Hong et al. \cite{hong2023watch} have designed an explicit scoring module to identify reliable streams and effectively manage input scenarios. Regarding the issue of missing video frames, most researchers have applied dropout techniques to fortify network robustness \cite{13:chang2022robustness,12:makino2019recurrent,9:shi2022learning,8:hazarika2022analyzing,11:zhang2019robust}. In classical dropout methods, per-frame dropout is initially utilized in \cite{11:zhang2019robust} and utterance-level dropout is applied in AV-Hubert \cite{9:shi2022learning}. As a recent work focusing on this issue, Chang et al. \cite{13:chang2022robustness} unify test suites of missing videos. However, the proposed binary evaluation metric overly emphasizes relative robustness trends, neglecting absolute performance. Compared to the methods mentioned earlier, we reassess the problem of missing video frames from the perspective of modality bias. Leveraging classical techniques and simple designs, our approach achieves both performance and robustness without introducing additional inference time. It adapts to various scenarios of frame absence through a unified model.

\vspace{-2pt}
\section{Conclusion}
In this work, we discover and analyze the essence of dropout-induced modality bias. Based on these findings, we proposed MBH to provide a systematic description of the relationship between modality bias and missing robustness in multimodal systems. Consequently, we propose a new multimodal distribution approximation with knowledge distillation approach to deal with missing video frames for AVSR. Furthermore, we apply adapters to handle videos with both partial and severe missing rates. For future work, we intend to validate our findings in this study across a wide range of multimodal applications beyond AVSR.

\bigskip
{
   \small
    \bibliographystyle{unsrt} 
    \bibliography{main}

\begin{thebibliography}{10}

\bibitem{23:ma2021end}
Pingchuan Ma, Stavros Petridis, and Maja Pantic.
\newblock End-to-end audio-visual speech recognition with conformers.
\newblock In {\em ICASSP 2021-2021 IEEE International Conference on Acoustics, Speech and Signal Processing (ICASSP)}, pages 7613--7617. IEEE, 2021.

\bibitem{4:pan2022leveraging}
Xichen Pan, Peiyu Chen, Yichen Gong, Helong Zhou, Xinbing Wang, and Zhouhan Lin.
\newblock Leveraging unimodal self-supervised learning for multimodal audio-visual speech recognition, 2022.

\bibitem{chen2023leveraging}
Chen Chen, Yuchen Hu, Qiang Zhang, Heqing Zou, Beier Zhu, and Eng~Siong Chng.
\newblock Leveraging modality-specific representations for audio-visual speech recognition via reinforcement learning.
\newblock In {\em Proceedings of the AAAI Conference on Artificial Intelligence}, volume~37, pages 12607--12615, 2023.

\bibitem{xu2020discriminative}
Bo~Xu, Cheng Lu, Yandong Guo, and Jacob Wang.
\newblock Discriminative multi-modality speech recognition.
\newblock In {\em Proceedings of the IEEE/CVF conference on Computer Vision and Pattern Recognition}, pages 14433--14442, 2020.

\bibitem{yu2020audio}
Jianwei Yu, Shi-Xiong Zhang, Jian Wu, Shahram Ghorbani, Bo~Wu, Shiyin Kang, Shansong Liu, Xunying Liu, Helen Meng, and Dong Yu.
\newblock Audio-visual recognition of overlapped speech for the lrs2 dataset.
\newblock In {\em ICASSP 2020-2020 IEEE International Conference on Acoustics, Speech and Signal Processing (ICASSP)}, pages 6984--6988. IEEE, 2020.

\bibitem{hong2022visual}
Joanna Hong, Minsu Kim, Daehun Yoo, and Yong~Man Ro.
\newblock Visual context-driven audio feature enhancement for robust end-to-end audio-visual speech recognition.
\newblock {\em arXiv preprint arXiv:2207.06020}, 2022.

\bibitem{haliassos2022jointly}
Alexandros Haliassos, Pingchuan Ma, Rodrigo Mira, Stavros Petridis, and Maja Pantic.
\newblock Jointly learning visual and auditory speech representations from raw data.
\newblock {\em arXiv preprint arXiv:2212.06246}, 2022.

\bibitem{ma2023auto}
Pingchuan Ma, Alexandros Haliassos, Adriana Fernandez-Lopez, Honglie Chen, Stavros Petridis, and Maja Pantic.
\newblock {Auto-AVSR: Audio-visual speech recognition with automatic labels}.
\newblock In {\em ICASSP 2023-2023 IEEE International Conference on Acoustics, Speech and Signal Processing (ICASSP)}, pages 1--5. IEEE, 2023.

\bibitem{sterpu2018attention}
George Sterpu, Christian Saam, and Naomi Harte.
\newblock Attention-based audio-visual fusion for robust automatic speech recognition.
\newblock In {\em Proceedings of the 20th ACM International conference on Multimodal Interaction}, pages 111--115, 2018.

\bibitem{sterpu2020teach}
George Sterpu, Christian Saam, and Naomi Harte.
\newblock {How to teach DNNs to pay attention to the visual modality in speech recognition}.
\newblock {\em IEEE/ACM Transactions on Audio, Speech, and Language Processing}, 28:1052--1064, 2020.

\bibitem{hu2023cross}
Yuchen Hu, Ruizhe Li, Chen Chen, Heqing Zou, Qiushi Zhu, and Eng~Siong Chng.
\newblock {Cross-Modal Global Interaction and Local Alignment for Audio-Visual Speech Recognition}.
\newblock {\em arXiv preprint arXiv:2305.09212}, 2023.

\bibitem{son2017lip}
Joon Son~Chung, Andrew Senior, Oriol Vinyals, and Andrew Zisserman.
\newblock Lip reading sentences in the wild.
\newblock In {\em Proceedings of the IEEE conference on computer vision and pattern recognition}, pages 6447--6456, 2017.

\bibitem{24:afouras2018deep}
Triantafyllos Afouras, Joon~Son Chung, Andrew Senior, Oriol Vinyals, and Andrew Zisserman.
\newblock Deep audio-visual speech recognition.
\newblock {\em IEEE transactions on pattern analysis and machine intelligence}, 44(12):8717--8727, 2018.

\bibitem{afouras2018lrs3}
Triantafyllos Afouras, Joon~Son Chung, and Andrew Zisserman.
\newblock {LRS3-TED}: a large-scale dataset for visual speech recognition.
\newblock {\em arXiv preprint arXiv:1809.00496}, 2018.

\bibitem{hong2023watch}
Joanna Hong, Minsu Kim, Jeongsoo Choi, and Yong~Man Ro.
\newblock {Watch or Listen: Robust Audio-Visual Speech Recognition with Visual Corruption Modeling and Reliability Scoring}.
\newblock In {\em Proceedings of the IEEE/CVF Conference on Computer Vision and Pattern Recognition}, pages 18783--18794, 2023.

\bibitem{Dai2023icme}
Yusheng Dai, Hang Chen, Jun Du, Xiaofei Ding, Ning Ding, Feijun Jiang, and Chin-Hui Lee.
\newblock {Improving Audio-Visual Speech Recognition by Lip-Subword Correlation Based Visual Pre-training and Cross-Modal Fusion Encoder}.
\newblock In {\em 2023 IEEE International Conference on Multimedia and Expo (ICME)}, pages 2627--2632. IEEE, 2023.

\bibitem{13:chang2022robustness}
Oscar Chang, Otavio de Pinho~Forin Braga, Hank Liao, Dmitriy~Dima Serdyuk, and Olivier Siohan.
\newblock On robustness to missing video for audiovisual speech recognition.
\newblock {\em Transactions on Machine Learning Research (TMLR)}, 2022.

\bibitem{12:makino2019recurrent}
Takaki Makino, Hank Liao, Yannis Assael, Brendan Shillingford, Basilio Garcia, Otavio Braga, and Olivier Siohan.
\newblock Recurrent neural network transducer for audio-visual speech recognition.
\newblock In {\em 2019 IEEE automatic speech recognition and understanding workshop (ASRU)}, pages 905--912. IEEE, 2019.

\bibitem{9:shi2022learning}
Bowen Shi, Wei-Ning Hsu, Kushal Lakhotia, and Abdelrahman Mohamed.
\newblock Learning audio-visual speech representation by masked multimodal cluster prediction, 2022.

\bibitem{8:hazarika2022analyzing}
Devamanyu Hazarika, Yingting Li, Bo~Cheng, Shuai Zhao, Roger Zimmermann, and Soujanya Poria.
\newblock Analyzing modality robustness in multimodal sentiment analysis, 2022.

\bibitem{11:zhang2019robust}
Shiliang Zhang, Ming Lei, Bin Ma, and Lei Xie.
\newblock Robust audio-visual speech recognition using bimodal {DFSMN} with multi-condition training and dropout regularization.
\newblock In {\em ICASSP 2019-2019 IEEE international conference on acoustics, speech and signal processing (ICASSP)}, pages 6570--6574. IEEE, 2019.

\bibitem{2:zhao2021missing}
Jinming Zhao, Ruichen Li, and Qin Jin.
\newblock Missing modality imagination network for emotion recognition with uncertain missing modalities.
\newblock In {\em Proceedings of the 59th Annual Meeting of the Association for Computational Linguistics and the 11th International Joint Conference on Natural Language Processing (Volume 1: Long Papers)}, pages 2608--2618, 2021.

\bibitem{6:ma2022multimodal}
Mengmeng Ma, Jian Ren, Long Zhao, Davide Testuggine, and Xi~Peng.
\newblock Are multimodal transformers robust to missing modality?
\newblock In {\em Proceedings of the IEEE/CVF Conference on Computer Vision and Pattern Recognition}, pages 18177--18186, 2022.

\bibitem{18:xue2022modality}
Zihui Xue, Zhengqi Gao, Sucheng Ren, and Hang Zhao.
\newblock The modality focusing hypothesis: {T}owards understanding crossmodal knowledge distillation, 2022.

\bibitem{chen2021correlating}
Hang Chen, Jun Du, Yu~Hu, Li-Rong Dai, Bao-Cai Yin, and Chin-Hui Lee.
\newblock Correlating subword articulation with lip shapes for embedding aware audio-visual speech enhancement.
\newblock {\em Neural Networks}, 143:171--182, 2021.

\bibitem{10:zhou2019modality}
Pan Zhou, Wenwen Yang, Wei Chen, Yanfeng Wang, and Jia Jia.
\newblock Modality attention for end-to-end audio-visual speech recognition.
\newblock In {\em ICASSP 2019-2019 IEEE International Conference on Acoustics, Speech and Signal Processing (ICASSP)}, pages 6565--6569. IEEE, 2019.

\bibitem{chen2022dearkd}
Xianing Chen, Qiong Cao, Yujie Zhong, Jing Zhang, Shenghua Gao, and Dacheng Tao.
\newblock Dearkd: data-efficient early knowledge distillation for vision transformers.
\newblock In {\em Proceedings of the IEEE/CVF Conference on Computer Vision and Pattern Recognition}, pages 12052--12062, 2022.

\bibitem{valverde2021there}
Francisco~Rivera Valverde, Juana~Valeria Hurtado, and Abhinav Valada.
\newblock There is more than meets the eye: Self-supervised multi-object detection and tracking with sound by distilling multimodal knowledge.
\newblock In {\em Proceedings of the IEEE/CVF Conference on Computer Vision and Pattern Recognition}, pages 11612--11621, 2021.

\bibitem{xue2021multimodal}
Zihui Xue, Sucheng Ren, Zhengqi Gao, and Hang Zhao.
\newblock Multimodal knowledge expansion.
\newblock In {\em Proceedings of the IEEE/CVF International Conference on Computer Vision}, pages 854--863, 2021.

\bibitem{peng2019correlation}
Baoyun Peng, Xiao Jin, Jiaheng Liu, Dongsheng Li, Yichao Wu, Yu~Liu, Shunfeng Zhou, and Zhaoning Zhang.
\newblock Correlation congruence for knowledge distillation.
\newblock In {\em Proceedings of the IEEE/CVF International Conference on Computer Vision}, pages 5007--5016, 2019.

\bibitem{zhang2021tip}
Renrui Zhang, Rongyao Fang, Wei Zhang, Peng Gao, Kunchang Li, Jifeng Dai, Yu~Qiao, and Hongsheng Li.
\newblock Tip-adapter: Training-free clip-adapter for better vision-language modeling.
\newblock {\em arXiv preprint arXiv:2111.03930}, 2021.

\bibitem{houlsby2019parameter}
Neil Houlsby, Andrei Giurgiu, Stanislaw Jastrzebski, Bruna Morrone, Quentin De~Laroussilhe, Andrea Gesmundo, Mona Attariyan, and Sylvain Gelly.
\newblock Parameter-efficient transfer learning for nlp.
\newblock In {\em International Conference on Machine Learning}, pages 2790--2799. PMLR, 2019.

\bibitem{rebuffi2017learning}
Sylvestre-Alvise Rebuffi, Hakan Bilen, and Andrea Vedaldi.
\newblock Learning multiple visual domains with residual adapters.
\newblock {\em Advances in neural information processing systems}, 30, 2017.

\bibitem{gao2023clip}
Peng Gao, Shijie Geng, Renrui Zhang, Teli Ma, Rongyao Fang, Yongfeng Zhang, Hongsheng Li, and Yu~Qiao.
\newblock Clip-adapter: Better vision-language models with feature adapters.
\newblock {\em International Journal of Computer Vision}, pages 1--15, 2023.

\bibitem{hu2021lora}
Edward~J Hu, Yelong Shen, Phillip Wallis, Zeyuan Allen-Zhu, Yuanzhi Li, Shean Wang, Lu~Wang, and Weizhu Chen.
\newblock Lora: {L}ow-rank adaptation of large language models, 2021.

\bibitem{9746683}
Hang Chen, Hengshun Zhou, Jun Du, Chin-Hui Lee, Jingdong Chen, Shinji Watanabe, Sabato~Marco Siniscalchi, Odette Scharenborg, Di-Yuan Liu, Bao-Cai Yin, Jia Pan, Jian-Qing Gao, and Cong Liu.
\newblock {The First Multimodal Information Based Speech Processing (Misp) Challenge: Data, Tasks, Baselines And Results}.
\newblock In {\em ICASSP 2022 - 2022 IEEE International Conference on Acoustics, Speech and Signal Processing (ICASSP)}, pages 9266--9270, 2022.

\bibitem{wang2023multimodal}
Zhe Wang, Shilong Wu, Hang Chen, Mao-Kui He, Jun Du, Chin-Hui Lee, Jingdong Chen, Shinji Watanabe, Sabato Siniscalchi, Odette Scharenborg, et~al.
\newblock The multimodal information based speech processing (misp) 2022 challenge: Audio-visual diarization and recognition.
\newblock In {\em ICASSP 2023-2023 IEEE International Conference on Acoustics, Speech and Signal Processing (ICASSP)}, pages 1--5. IEEE, 2023.

\bibitem{lee2021intermediate}
Jaesong Lee and Shinji Watanabe.
\newblock Intermediate loss regularization for ctc-based speech recognition.
\newblock In {\em ICASSP 2021-2021 IEEE International Conference on Acoustics, Speech and Signal Processing (ICASSP)}, pages 6224--6228. IEEE, 2021.

\bibitem{wang2022sjtu}
Wei Wang, Xun Gong, Yifei Wu, Zhikai Zhou, Chenda Li, Wangyou Zhang, Bing Han, and Yanmin Qian.
\newblock The sjtu system for multimodal information based speech processing challenge 2021.
\newblock In {\em ICASSP 2022-2022 IEEE International Conference on Acoustics, Speech and Signal Processing (ICASSP)}, pages 9261--9265. IEEE, 2022.

\bibitem{xu2022channel}
Gaopeng Xu, Song Yang, Wei Li, et~al.
\newblock {Channel-Wise AV-Fusion Attention for Multi-Channel Audio-Visual Speech Recognition}.
\newblock In {\em Proc. ICASSP 2022}, pages 9251--9255. IEEE, 2022.

\bibitem{yang2019lrw}
Shuang Yang, Yuanhang Zhang, Dalu Feng, Mingmin Yang, Chenhao Wang, Jingyun Xiao, Keyu Long, Shiguang Shan, and Xilin Chen.
\newblock Lrw-1000: A naturally-distributed large-scale benchmark for lip reading in the wild.
\newblock In {\em 2019 14th IEEE international conference on automatic face \& gesture recognition (FG 2019)}, pages 1--8. IEEE, 2019.

\bibitem{nio2022}
Sang~Wang Gaopeng~Xu, Xianliang~Wang et~al.
\newblock The {NIO} system for audio-visual diarization and recognition in {MISP} challenge 2022.
\newblock \url{https://mispchallenge.github.io/mispchallenge2022/papers/task2/Track2_NIO.pdf}, 2022.

\bibitem{li2023xmu}
Tao Li, Haodong Zhou, Jie Wang, Qingyang Hong, and Lin Li.
\newblock {The XMU System for Audio-Visual Diarization and Recognition in MISP Challenge 2022}.
\newblock In {\em ICASSP 2023-2023 IEEE International Conference on Acoustics, Speech and Signal Processing (ICASSP)}, pages 1--2. IEEE, 2023.

\bibitem{wang2024mlca}
He~Wang, Pengcheng Guo, Pan Zhou, and Lei Xie.
\newblock Mlca-avsr: Multi-layer cross attention fusion based audio-visual speech recognition.
\newblock {\em arXiv preprint arXiv:2401.03424}, 2024.

\bibitem{watanabe2020chime}
Shinji Watanabe, Michael Mandel, Jon Barker, Emmanuel Vincent, Ashish Arora, Xuankai Chang, Sanjeev Khudanpur, Vimal Manohar, Daniel Povey, Desh Raj, et~al.
\newblock Chime-6 challenge: Tackling multispeaker speech recognition for unsegmented recordings.
\newblock {\em arXiv preprint arXiv:2004.09249}, 2020.

\bibitem{guo2023npu}
Pengcheng Guo, He~Wang, Bingshen Mu, Ao~Zhang, and Peikun Chen.
\newblock {The NPU-ASLP System for Audio-Visual Speech Recognition in MISP 2022 Challenge}.
\newblock In {\em ICASSP 2023-2023 IEEE International Conference on Acoustics, Speech and Signal Processing (ICASSP)}, pages 1--2. IEEE, 2023.

\bibitem{fiscus1997post}
Jonathan~G Fiscus.
\newblock A post-processing system to yield reduced word error rates: {R}ecognizer output voting error reduction ({ROVER}).
\newblock In {\em Proc. asrU 1997}, pages 347--354. IEEE, 1997.

\bibitem{cheng2023whu}
Ming Cheng, Haoxu Wang, Ziteng Wang, Qiang Fu, and Ming Li.
\newblock The whu-alibaba audio-visual speaker diarization system for the misp 2022 challenge.
\newblock In {\em ICASSP 2023-2023 IEEE International Conference on Acoustics, Speech and Signal Processing (ICASSP)}, pages 1--2. IEEE, 2023.

\bibitem{19:suo2019metric}
Qiuling Suo, Weida Zhong, Fenglong Ma, Ye~Yuan, Jing Gao, and Aidong Zhang.
\newblock {Metric Learning on Healthcare Data with Incomplete Modalities.}
\newblock In {\em IJCAI}, volume 3534, page 3540, 2019.

\bibitem{20:cai2018deep}
Lei Cai, Zhengyang Wang, Hongyang Gao, Dinggang Shen, and Shuiwang Ji.
\newblock Deep adversarial learning for multi-modality missing data completion.
\newblock In {\em Proceedings of the 24th ACM SIGKDD international conference on knowledge discovery \& data mining}, pages 1158--1166, 2018.

\bibitem{21:wang2020transmodality}
Zilong Wang, Zhaohong Wan, and Xiaojun Wan.
\newblock Transmodality: {A}n end2end fusion method with transformer for multimodal sentiment analysis.
\newblock In {\em Proceedings of The Web Conference 2020}, pages 2514--2520, 2020.

\bibitem{22:pham2019found}
Hai Pham, Paul~Pu Liang, Thomas Manzini, Louis-Philippe Morency, and Barnab{\'a}s P{\'o}czos.
\newblock Found in translation: {L}earning robust joint representations by cyclic translations between modalities.
\newblock In {\em Proceedings of the AAAI Conference on Artificial Intelligence}, volume~33, pages 6892--6899, 2019.

\bibitem{li2023mseg3d}
Jiale Li, Hang Dai, Hao Han, and Yong Ding.
\newblock Mseg3d: Multi-modal 3d semantic segmentation for autonomous driving.
\newblock In {\em Proceedings of the IEEE/CVF Conference on Computer Vision and Pattern Recognition}, pages 21694--21704, 2023.

\bibitem{varga1993noisex}
A~Varga, HJM Steeneken, et~al.
\newblock Noisex-92: A database and an experiment to study the effect of additive noise on speech recognition systems.
\newblock {\em Speech Commun}, 12(3):247--253, 1993.

\bibitem{drude2018nara}
Lukas Drude, Jahn Heymann, Christoph Boeddeker, and Reinhold Haeb-Umbach.
\newblock Nara-wpe: A python package for weighted prediction error dereverberation in numpy and tensorflow for online and offline processing.
\newblock In {\em Speech Communication; 13th ITG-Symposium}, pages 1--5. VDE, 2018.

\bibitem{boeddecker18_chime}
Christoph Boeddecker, Jens Heitkaemper, Joerg Schmalenstroeer, et~al.
\newblock {Front-end processing for the CHiME-5 dinner party scenario}.
\newblock In {\em Proc. CHiME 2018}, pages 35--40, 2018.

\bibitem{raj2022gpu}
Desh Raj, Daniel Povey, and Sanjeev Khudanpur.
\newblock {GPU}-accelerated guided source separation for meeting transcription, 2022.

\bibitem{chen2022audio}
Hang Chen, Jun Du, Yusheng Dai, Chin~Hui Lee, Sabato~Marco Siniscalchi, Shinji Watanabe, Odette Scharenborg, Jingdong Chen, Bao~Cai Yin, and Jia Pan.
\newblock Audio-visual speech recognition in misp2021 challenge: {D}ataset release and deep analysis.
\newblock In {\em Proceedings of the Annual Conference of the International Speech Communication Association, INTERSPEECH}, volume 2022, pages 1766--1770, 2022.

\bibitem{1:ma2021smil}
Mengmeng Ma, Jian Ren, Long Zhao, Sergey Tulyakov, Cathy Wu, and Xi~Peng.
\newblock {Smil: Multimodal learning with severely missing modality}.
\newblock In {\em Proceedings of the AAAI Conference on Artificial Intelligence}, volume~35, pages 2302--2310, 2021.

\bibitem{16:gat2020removing}
Itai Gat, Idan Schwartz, Alexander Schwing, and Tamir Hazan.
\newblock Removing bias in multi-modal classifiers: {R}egularization by maximizing functional entropies.
\newblock {\em Advances in Neural Information Processing Systems}, 33:3197--3208, 2020.

\bibitem{17:johnson2017clevr}
Justin Johnson, Bharath Hariharan, Laurens Van Der~Maaten, Li~Fei-Fei, C~Lawrence~Zitnick, and Ross Girshick.
\newblock Clevr: {A} diagnostic dataset for compositional language and elementary visual reasoning.
\newblock In {\em Proceedings of the IEEE conference on computer vision and pattern recognition}, pages 2901--2910, 2017.

\bibitem{7:guo2023modality}
Yangyang Guo, Liqiang Nie, Harry Cheng, Zhiyong Cheng, Mohan Kankanhalli, and Alberto Del~Bimbo.
\newblock On modality bias recognition and reduction.
\newblock {\em ACM Transactions on Multimedia Computing, Communications and Applications}, 19(3):1--22, 2023.

\end{thebibliography}
}

\clearpage
\setcounter{page}{1}
\maketitlesupplementary

\section {Additional Experiments}

\paragraph{Analysis of Latent Distribution Samples} We further analyze the latent distribution samples of the proposed robust AVSR model to validate two conclusions: (1) In Figure \ref{fig:emb1}, our goal is to prove that MDA-KD effectively avoids dropout-induced bias and compels the model to employ a collaborative decision strategy, even with video frames missing input. (2) In Figure \ref{fig:emb2}, we aim to demonstrate that activating the MS-Adapter indeed dynamically switches the decision-making pattern to an audio-dominant one when facing complete video missing input.

\paragraph{Analysis of Zero-shot Noise Robustness} We further evaluate the system performance with zero-shot noise. Specifically, we introduce Babble noise from NOISEX \cite{varga1993noisex}, unseen during training, into the near-field audio captured by a head-worn microphone at a specific SNR level. In Table \ref{table2:drpout comparison}, we reused the symbols from Table \ref{table2:drpout comparison}. Our results demonstrate that our proposed modality-unbiased model, AV6, outperforms both the modality-biased model AV1 and the unimodal model A0 in both Near Field and Far Field settings with in-set noise. More importantly, the advantage is further highlighted with zero-shot noise across all SNR levels, aligning with the objective of AVSR as a robust system for real-world applications.


\paragraph{Analysis of Computational Consumption} \vskip -10pt
We present an analysis of computational consumption measured in FLOPS when faced with complete video absence, offering insights into the effective reduction achieved upon activating the MS-Adapter by interrupting the data flow in the video branch. When activating MS-Adapter, data exclusively flows through the audio branch, involving a mere 3.89 GFLOPS and 94.21 M parameters in a computation. This stands in favorable contrast to conventional methods that require padding video tensor inputs, demanding 12.64 GFLOPS and totaling 144.78 M parameters.

\paragraph{Experiment Details on Different Test Dropout Methods} \vskip -10pt
In Figure \ref{fig:singel}, we provide more comprehensive experimental results and present performance degradation curves across all three test suites (Segment Dropout, Utterance Dropout, and Interval Dropout) to facilitate further research.

 \captionsetup[table]{belowskip=-10pt}
\begin{table}[!t]
\resizebox{1.0\columnwidth}{!}{
\begin{tabular}{cccccc}\toprule[1pt]
\multirow{2}{*}{\textbf{Models}} & \multirow{2}{*}{\textbf{Near Field}} & \multirow{2}{*}{\textbf{Far Field}} & \multicolumn{3}{c}{\textbf{Zero-shot Babble Noise}} \\ \cline{4-6} 
                                 &                                      &                                     & 0dB             & -2.5dB           & -5dB           \\ \hline
A0                               & 18.10                                & 25.13                               & 33.52           & 62.17            & 75.76          \\
AV1                              & 17.71                                & 23.26                               & 29.40           & 51.63            & 63.80          \\
AV6                              & \textbf{16.86}                       & \textbf{21.11}                      & \textbf{26.67}  & \textbf{44.97}   & \textbf{55.65} \\ \bottomrule[1pt]
\end{tabular}}
\caption{CER comparison of zero-shot noise roubustness.}
\label{table:zero-shot}
\end{table}

\section{Distinctions between MBVD and MVD}
There are three key distinctions between the Modality Bias Venn Diagram (MBVD) and the Modality Venn Diagram (MVD) \cite{18:xue2022modality}. Firstly, MBVD focuses on the hidden feature space to describe the decision pattern of a multimodal model, while MVD space is essentially another form of the original feature space. Secondly, for the generation order, MBVD maps from the original feature space $\mathcal{X}$ to the decisive feature space $\mathcal{Z}$, while MVD follows the opposite direction. Lastly, MBVD is employed to describe modality bias in decision-making processes, whereas MVD is utilized for knowledge distillation.

\begin{figure}[!t]
    \centering
     \includegraphics[width=1.0\columnwidth]{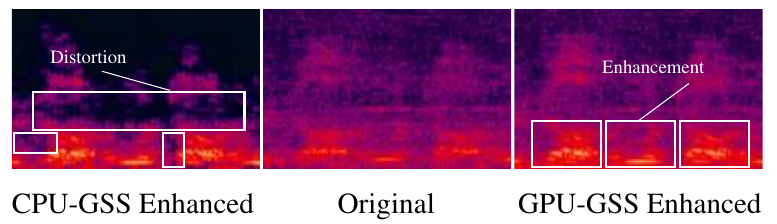}
    \caption{Spectral analysis of GSS-enhanced signals}
    \label{fig:spectrum}
\end{figure}
\begin{figure*}[!t]
    \centering
     \includegraphics[width=1.85\columnwidth]{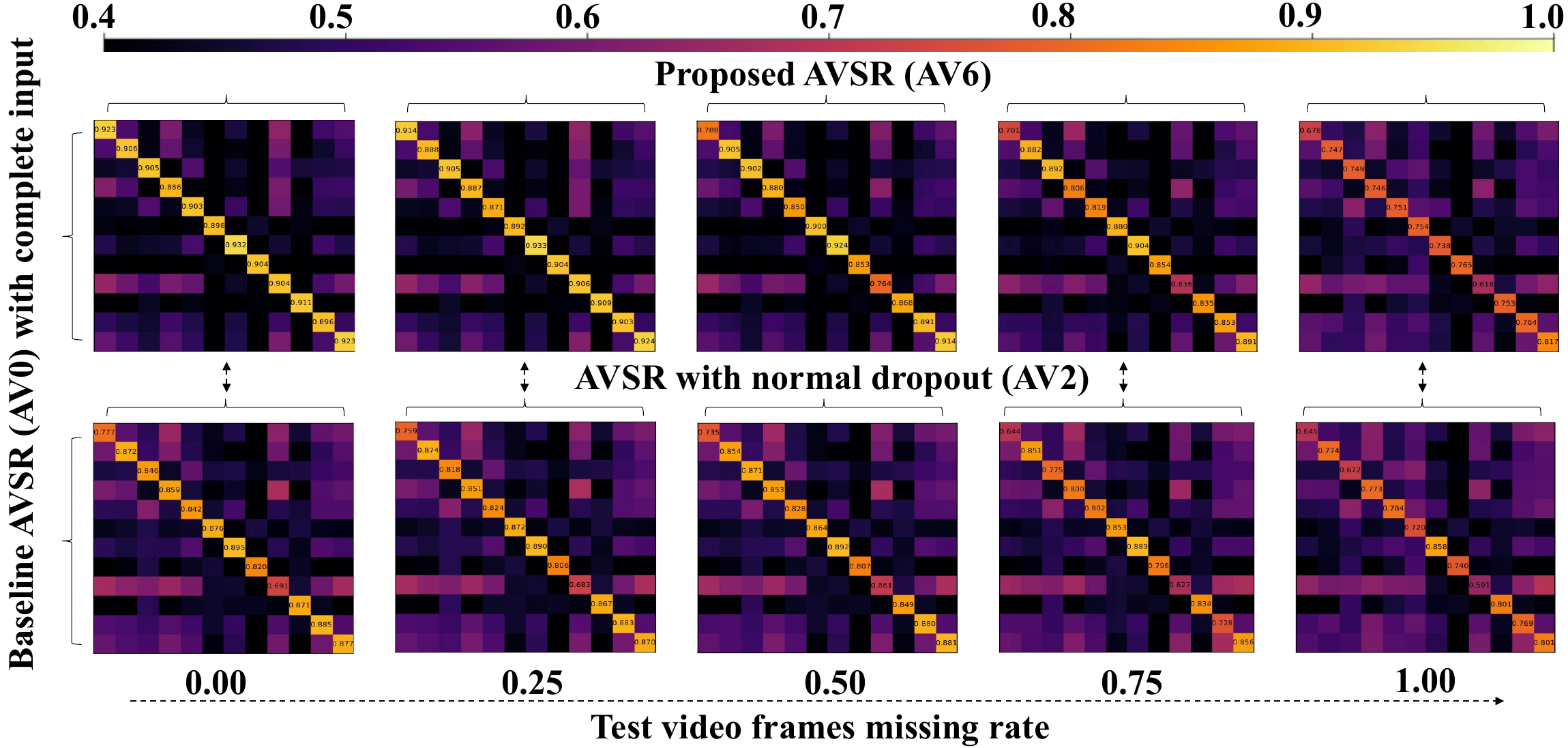}
    \caption{We investigate the decision discrepancies between the proposed robust AVSR (AV6) and the AVSR trained using the normal dropout technique (AV2) across different test video frames missing rates. Similar to Figure \ref{fig3:Training_Droprate}, we quantify the divergence by calculating the cosine distance similarity of latent decision distribution samples from both models with missing video frames input and those of AV0 with complete data input. The latter samples represent an ideal collaborative decision strategy. Each diagonal element in the cosine distance-based similarity matrix represents the similarity between intermediate representations with the same sample index but may have different missing rates. As a result, two prominent phenomena emerge. (1) In vertical comparison between AV6 and AV2, the sample similarities of AV6 consistently surpass those of AV2 along the diagonal line, indicating a closer approximation to the ideal collaborative decision distribution in latent space. These results suggest that MDA-KD enables AV6 to adopt a decision strategy similar to AV0, whether facing complete input or missing video frames, effectively utilizing content information and modality general information audio modality. (2) In horizontal comparison, in the first row, the diagonal elements in each subplot consistently darken as the missing rate increases, and the last subplot darkens sharply with the shift of decisive bias on audio modality upon activating the MS-Adapter. This trend is less pronounced in the second row, as AV2 exhibits an excessive modality bias on audio modality, deviating from the collaborative decision strategy.}
    \label{fig:emb1}
\end{figure*}

\begin{figure*}[!t]
    \centering
     \includegraphics[width=1.85\columnwidth]{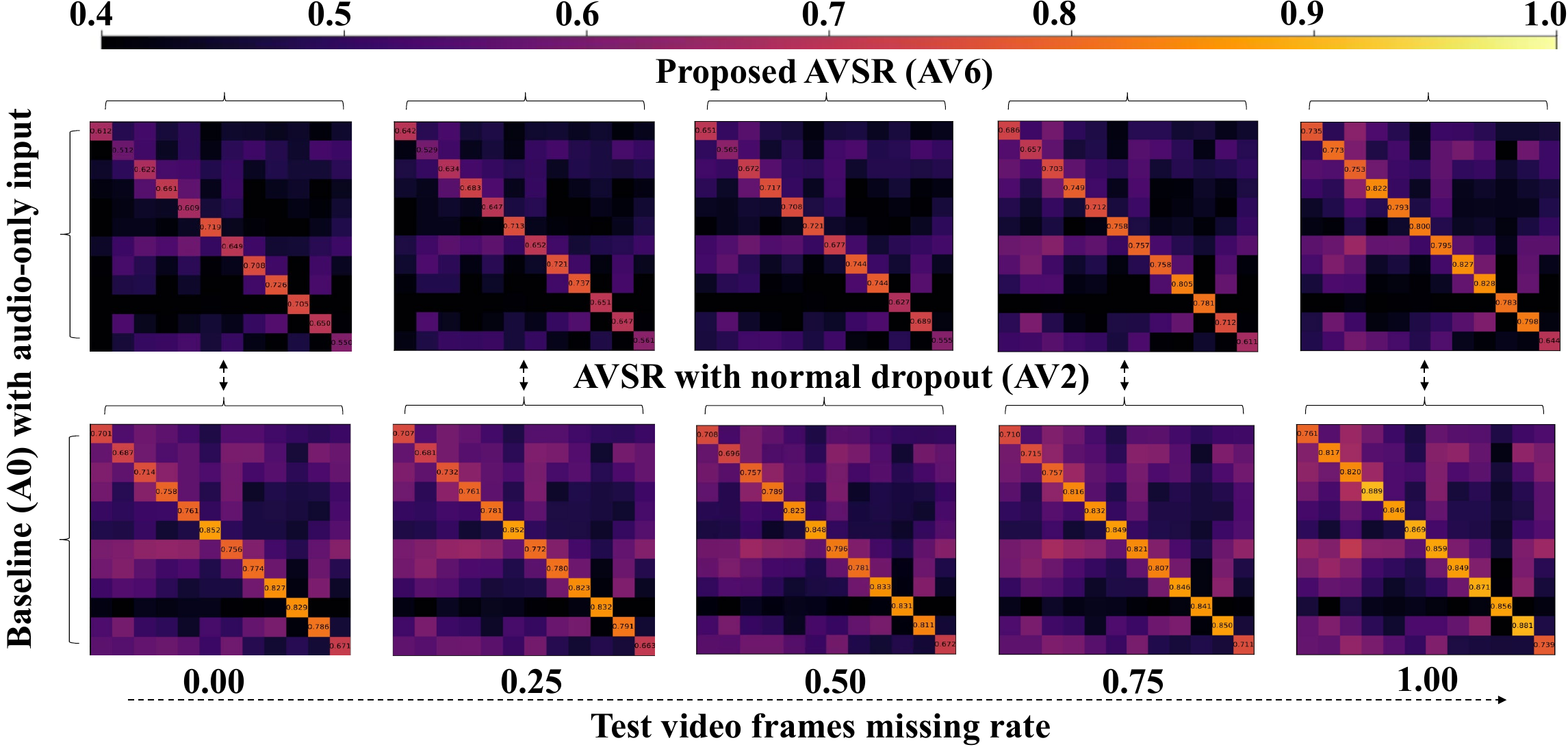}
    \caption{We compare the decision discrepancies between AV6 and AV2 with A0, revealing two distinct phenomena. (1) In the first row, the diagonal line of the last subplot sharply brightens compared to the former four subplots, indicating the effectiveness of the MS-Adapter in dynamically switching the decisive pattern towards the audio-dominant one. (2) In comparison to the first row, the diagonal line of the second row remains consistently bright across various missing video frame rate inputs. This further confirms that AV2 is a modality-biased model that consistently relies on the audio modality.}
    \label{fig:emb2}
\end{figure*}
\begin{figure*}[!t]
    \centering
     \includegraphics[width=2.0\columnwidth]{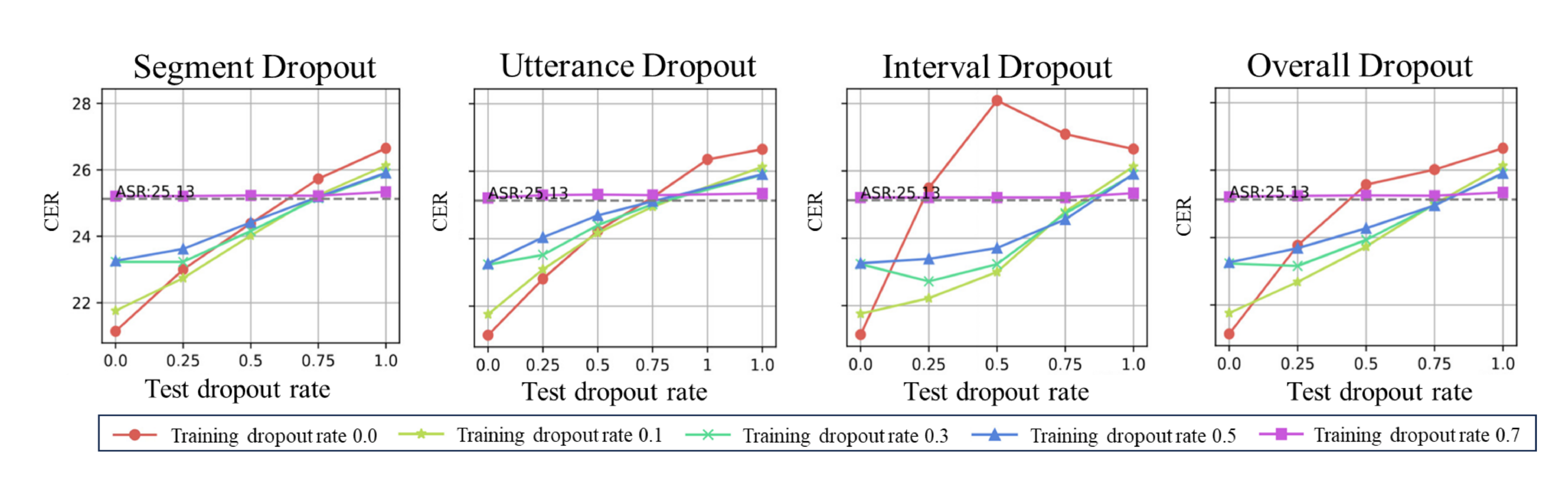}
    \caption{Performance degradation curves of AVSR systems with different training dropout rate test in different test dropout methods.}
    \label{fig:singel}
\end{figure*}

\section{Implement Details}
\paragraph{Data Processing Details}
We apply conventional signal processing algorithms, such as weighted prediction error (WPE) \cite{drude2018nara} and guided source separation (GSS) \cite{boeddecker18_chime}, to multichannel far/middle-field audio for dereverberation and source separation in both the training and test sets. Specifically, we utilize a GPU-accelerated version of GSS \cite{raj2022gpu}. As shown in Figure \ref{fig:spectrum}, it effectively enhances the spectral speech components for the target speaker while minimizing speech distortion compared to the CPU version \cite{boeddecker18_chime}. We then apply a short Fourier transform and Mel filtering to obtain 80-dimensional filterbank frames in the frequency domain, with a 0.25s window length and a 0.01s frameshift, using a 16k sample rate. For video, following~\cite{chen2022audio}, we acquire grayscale lip ROI with 88$\times$88 pixels before inputting it into the network. In ASR training, all enhanced far/near/middle-field audio is used, employing various data augmentation techniques, such as adding noise, Room Impulse Response (RIR) convolution, speed perturbation, and concatenating nearby segments to create a 10-fold training set. The technique of concatenating nearby segments effectively generates a longer segment, providing additional content information. This technique can used in both training and decoding phrases. For VSR, we pre-train the visual frontend on far/middle-filed video following \cite{Dai2023icme} by correlating lip shapes with syllabic HMM states (3168 Senone units). In AVSR training, the audio and visual branches are initialized with pre-trained ASR and VSR representations. We create an 8-fold training set, incorporating two effective data augmentation techniques: (1) matching synchronous audio and video segments recorded in different fields and (2) concatenating nearby segments in both video and audio.

\paragraph{Training Implementation Details}
All conformers in our network use the same set of hyperparameters ($n_{\text{head}} = 8$, $d_{\text{model}} = 512$, $d_{\text{ffn}} = 2048$, $CNN_{kernel} = 5$). The decoder consists of six transformer blocks ($n_{head} = 8$, $d_{\text{model}} = 512$, $d_{\text{ffn}} = 2048$). For unimodal model training, we strictly adhere to \cite{Dai2023icme}. In robustness training for this work, all models are optimized using Adam with $\beta_1 = 0.9$, $\beta_2=0.999$, and a learning rate of $0.0012$. For MDA-KD implementation in Section \ref{sec:experiment}, we utilize the intermediate representation samples from the output of ResNet-18 and the first layer of Conformer in the video branch in practice. For further exploration, we successfully validate that the output of the multimodal encoder exhibits similar effectiveness in achieve both missing robustness and accuracy with complete input. The learning rate undergoes a linear warm-up during the first 3000 steps and subsequently decreases proportionally to the inverse square root of the step number. We train for $12$ epochs with a training batch size of $128$, utilizing $4$ NVIDIA Tesla A100 48GB GPUs. For MS-Adapter adaptation, we train $5$ epochs with a batch size of $144$ and a learning rate of $0.0002$. During decoding, the beam size is set to $10$ in beam search. Additionally, a 6-layer transformer-based language model trained on the transcription of the training set is employed in decoding, with a weight of $0.2$, although it brings negligible performance improvement.

\paragraph{Dropout Setting Details}
Segment Dropout, Utterance Dropout, and Interval Dropout are employed to simulate missing video modality in different scenarios. Segment dropout occurs when contiguous segments of video frames are dropped, which often occurs when the lips are covered or when the person is in a side-face pose. Utterance Dropout refers to dropping the entire video, which represents situations where the camera is turned off. Interval Dropout means dropping ($dropout\ rate < 0.5$) or preserving ($dropout\ rate > 0.5$) video frames at a fixed interval, indicating missing due to network latency or hardware computation bottleneck. Unlike previous work \cite{13:chang2022robustness}, we have simplified the test suites by removing frame-level random dropout to ensure experimental reproducibility. Furthermore, the starting position for segment dropout is randomly determined. Considering our study on modality bias and robustness, the focus lies more on the dropout rate than the dropout method.

\section{More Discussions on Related Works}
\label{subsec:Missing Modality}
\paragraph{Missing Modality in Multimodal Learning} The missing modalities problem is common in multimodal applications,  whether in the training or testing stage, and has attracted a lot of research interest. For modality-balanced models like Multimodal Emotion Recognition (MER) and multimodal sensor fusion in autonomous driving, the mainstream approach is to learn joint multimodal representations to capture intra- or inter-modal features cross modalities \cite{21:wang2020transmodality,22:pham2019found}.  For modality-biased models, data augmentation methods such as modality dropout effectively address out-of-distribution issues  \cite{9:shi2022learning,12:makino2019recurrent,13:chang2022robustness}. In cases of  severe modality absence, generative models \cite{19:suo2019metric,20:cai2018deep} and meta-learning based methods  \cite{1:ma2021smil} are used to directly predict the missing modalities based on available modalities or a few-shot paired samples. For AVSR, we prioritize efficiency and opt for dropout due to its plug-and-play nature and lightweight implementation.

\paragraph{Modality Bias in Multimodal Learning}
The modality bias is observed in many multimodal applications, since there is a direct correlation between a specific modality and the target task, leading to one modality dominating the decision-making process \cite{16:gat2020removing}. In the VQA, several de-bias methods have been proposed. New datasets following the answer distribution balancing rule have been constructed to address the language prior problem \cite{17:johnson2017clevr}. Guo et al. \cite{7:guo2023modality} develop plug-and-play loss function methods that can adaptively learn the feature space for each label. Gat et al. \cite{16:gat2020removing} have proposed a method based on the log-Sobolev inequality. Although many studies have been conducted on removing bias, there is a lack of conception or mathematical models to describe model bias and limited research on the impact of bias on the modality missing problem.

\paragraph{Dropout-Induced Modality Bias on Mulitmodal Tasks}
For AVSR, this excessive modality bias towards audio is a double-edged sword, as it brings robustness to missing video data while degrading the performance of a multimodal model on complete multimodal data. It causes the model to tend towards trivial solutions and ignore optimal ones. As a result, the model neglects visual cues, making it sensitive to perturbations in speech. This contradicts the intention of AVSR as a multimodal robust speech recognition application in noisy environments.

For other multimodal applications, Hazarika et al. investigate the robustness of Multimodal Sentiment Analysis (MSA), which is a multimodal classifier with text, visual, and audio as input \cite{8:hazarika2022analyzing}. By applying dropout on the training text, the robustness against missing text can be achieved without compromising the original performance. These findings seem to be inconsistent with the degradation results observed in AVSR. While the truth is that the common MSA system exhibits a severe modality bias dominated by text, and it is sensitive to perturbations in text but robust to other modalities. Applying dropout on text helps to mitigate over-reliance and encourages the model to leverage supplementary information across modalities. A similar phenomenon has been observed in AVSR when applying dropout on the audio modality \cite{9:shi2022learning,12:makino2019recurrent}. Interestingly, in our research on video robustness in AVSR, video is a supplementary modality within the system rather than the dominant one. As a result, we emphasize that it is important to first determine whether the system has a dominant or supplementary modality when studying the robustness of a specific modality within a multimodal bias system.

\section{Limitations}
Modality dropout presents two facets. On one hand, it could address the out-of-distribution (OOD) issue resulting from missing modalities. On the other hand, if applied on supplementary modalities, it can induce dropout-induced modality bias in modality-biased systems. For our further exploration, we observe the manifestation of these characteristics is related to input quality. In this work, we focus on real-world TV room scenarios with relatively low-resolution video and noisy speech. Under such conditions, dropout-induced modality bias is observed prominently. While for high-quality datasets, such as LRS2 and LRS3, dropout serves more as a form of data augmentation, and the dropout-induced modality bias are mitigated by high-quality input. Nevertheless, in all conditions, the proposed MDA-KD and MS-Adapter consistently lead to relative improvements to original dropout method.


\end{document}